\documentclass[%
 aip,
 amsmath,amssymb,
 reprint,%
]{revtex4-2}

\usepackage{graphicx}
\graphicspath{{figures/}}
\draft 
\usepackage{amsmath}
\usepackage{float}
\usepackage{siunitx}
\makeatletter
\let\newfloat\newfloat@ltx
\makeatother

\begin{document}


\title{General kinetic ion induced electron emission model for metallic walls applied to biased Z-pinch electrodes}   



\author{C. R. Skolar}  
\email[]{chirag.skolar@njit.edu}
\affiliation{Center for Solar-Terrestrial Research, New Jersey Institute of Technology, Newark, NJ 07102, USA}

\author{K. Bradshaw}
\affiliation{Department of Astrophysical Sciences, Princeton University, Princeton, NJ 08544, USA}

\author{M. Francisquez}
\affiliation{Princeton Plasma Physics Laboratory, Princeton, NJ 08540, USA}

\author{L. Murillo}
\affiliation{William E. Boeing Department of Aeronautics and Astronautics, University of Washington, Seattle, WA 98195, USA}

\author{V. Krishna Kumar}
\affiliation{William E. Boeing Department of Aeronautics and Astronautics, University of Washington, Seattle, WA 98195, USA}

\author{B. Srinivasan}
\email[]{srinbhu@uw.edu}
\affiliation{William E. Boeing Department of Aeronautics and Astronautics, University of Washington, Seattle, WA 98195, USA}

\date{\today}

\begin{abstract}
	A kinetic ion induced electron emission (IIEE) model for general applications is developed to obtain the emitted electron energy spectrum for a distribution of ion impacts on a metallic surface.
	We assume an ionization cascade mechanism and use empirical models for the ion and electron stopping powers.
	The emission spectrum and the secondary electron yield (SEY) are validated for a variety of materials.
	The IIEE model is used to study the effect of IIEE on the plasma-material interactions of Z-pinch electrodes.
	Un-magnetized Boltzmann-Poisson simulations are performed for a Z-pinch plasma doubly bounded by two biased copper electrodes with and without IIEE at bias potentials from 0 to 9 kV.
	At the anode, the SEY decreases from 0 to 1 kV, but then increases at higher bias potentials.
	At the cathode, the SEY is much larger due to higher energy ion bombardment and grows with bias potential.
	As the bias potential increases, the emitted cathode electrons are accelerated to higher energies into the domain collisionally heating the plasma.
	Above 1 kV, the heating is strong enough to increase the plasma potential.
	Despite SEY greater than 1, only a classical sheath forms as opposed to a space-charge limited or inverse sheath due to the emitted electron flux not reaching the space charge current saturation limits.
	Furthermore, the current in the emissionless cases saturates to a value lower than experiment.
	With IIEE, the current does not saturate and continues to increase with the 4 kV case matching most closely with experiment.
\end{abstract}


\maketitle 


\section{Introduction}  
\label{s:intro}

Understanding the plasma-material interactions near plasma facing components is vital for a variety of applications including electric thrusters, plasma etching, Langmuir probes, spacecraft charging, and nuclear fusion reactor walls. \cite{stangeby2000plasma,lieberman2005principles} 
The wall material may emit secondary electrons due to electron impacts, ion impacts, high temperatures (thermionic emission), and photon impacts (photoemission).
These secondary electron emissions are important to model because they can change the behavior of the plasma density, current, and potential.
This work focuses on ion induced electron emission (IIEE).
We create a general framework to model IIEE kinetically by accounting for the energy spectrum of both the impacting ions and emitted electrons.

Ion induced electron emission (IIEE) becomes a relevant emission mechanism in the \si{\kilo\electronvolt} regime. 
Many previous plasma-material interaction studies have been at lower energy regimes and therefore have been able to neglect the effects of IIEE. 
For applications with high ion energies, IIEE is expected to play an important role.
In particular, this work applies the generalized IIEE framework to study the plasma-material interactions near the electrodes of Z-pinch fusion devices.

Shear flow stabilized Z-pinches are a promising path towards achieving commercial fusion energy. \cite{levitt2023}
They use two biased electrodes to generate a large axial current in the plasma.
This causes a self-generated azimuthal magnetic field that confines the plasma.
Increasing the current results in a larger confinement magnetic field strength which yields higher plasma densities and temperatures leading to a greater nuclear fusion reaction rate.
With sufficient scaling of the current, the shear flow stabilized Z-pinch is expected to achieve breakeven (more output energy than input energy) and beyond.\cite{shumlak2020z,levitt2023}
To push towards a commercially viable fusion pilot plant, it is important to understand how the naturally high Z-pinch particle and heat fluxes at the electrode interfaces\cite{skolar2023} may cause degradation.\cite{thompson2024}  

Near the electrode walls (and generally for all plasma facing components), the higher mobility electrons are more readily absorbed into the wall as compared to the ions.
This results in a potential drop that acts as a barrier for the electrons and accelerates the ions.
The result is a thin layer of positive space charge near the wall called the plasma sheath that is on the order of a few Debye lengths,  
$\lambda_D=\sqrt{\epsilon_0 T_e/ne^2}$, where  
$\epsilon_0$ is the permittivity of free space, 
$T_e$ is the electron temperature in energetic units, 
$n$ is the plasma density, 
and $e$ is the elementary electric charge.
In cases without an electric potential bias, there is no net current so the ion and electron particle fluxes are equal.
However, in the presence of an electric potential bias, as with Z-pinch electrodes, a current flows through the plasma-wall interface.\cite{stangeby2000plasma,baalrud2020interaction,skolar2023}

In the limit of a perfectly absorbing wall, the current saturates with increasing bias potential.\cite{stangeby2000plasma,skolar2023}
This is because the ion particle flux is negligibly affected by the bias potential and the perfectly absorbing wall assumption constrains the electron particle flux to either be zero or go into the wall.
Thus, as the bias potential increases, the electron particle flux tends to zero limiting the total current within the plasma.\cite{skolar2023}
While this is expected based on theory,\cite{stangeby2000plasma} the magnitude of the saturation current is lower\cite{skolar2023} than what has been measured in Z-pinch experiments.\cite{zhang2019sustained}
The difference may be explained by the inclusion of IIEE, which will allow electron particle flux out of the wall.\cite{stangeby2000plasma}

\begin{figure}[!htb]
	\centering
	\includegraphics[width=\linewidth]{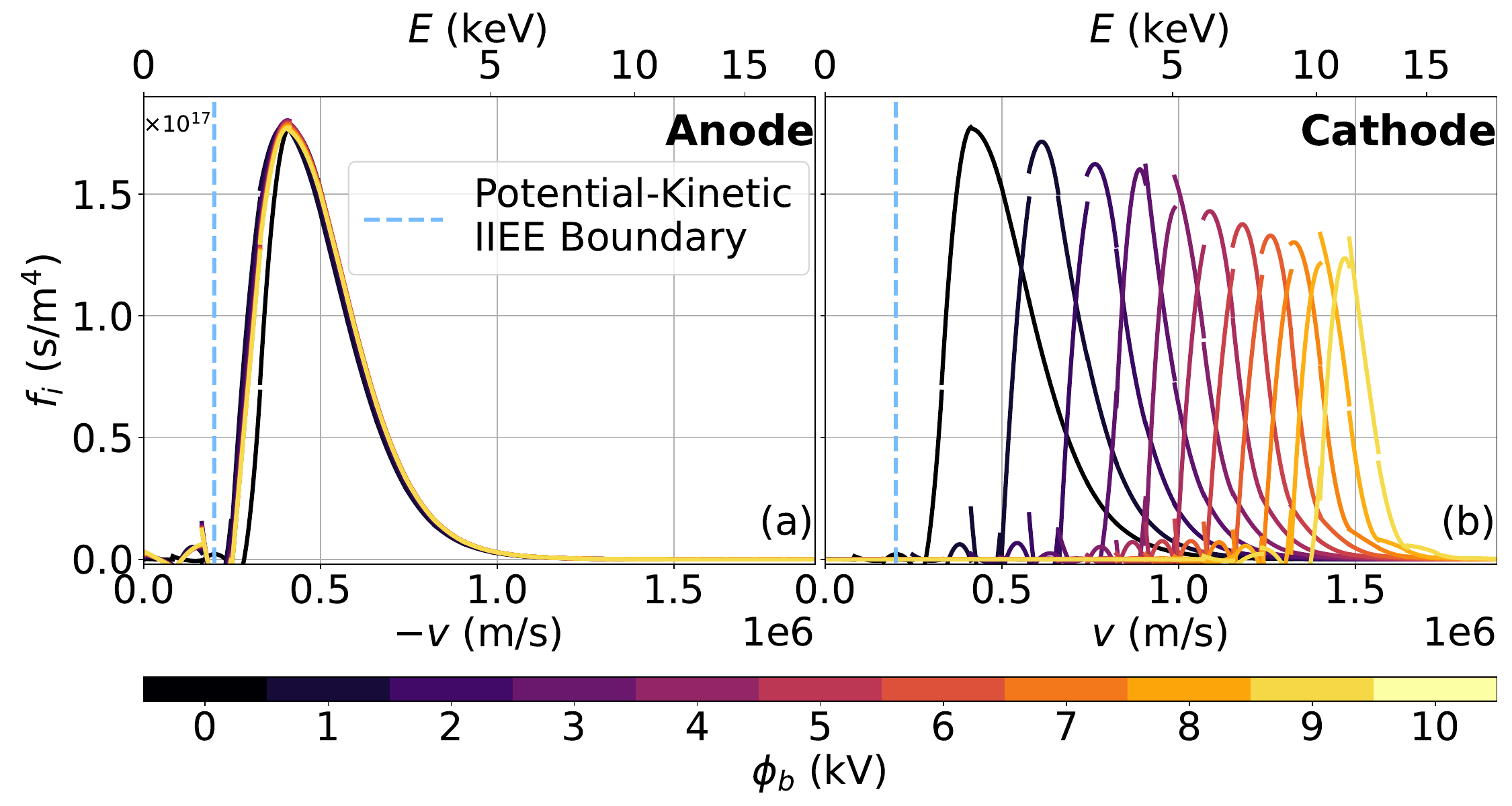}
	\caption{
		Plots of the ion distribution at the anode (a) and cathode (b) for electric potential biases ranging from \qtyrange{0}{10}{\kilo\volt}.\cite{skolar2023} 
		The anode is minimally affected by bias potential. 
		As the bias potential increases, while there are fewer ions at the cathode, they have much higher energies.
		The dashed vertical line corresponds to the boundary ($v=\SI[exponent-product=\times,output-exponent-marker=,per-mode=symbol]{2e5}{\meter\per\second}$) at which kinetic IIEE dominates over potential IIEE.\cite{benazeth1982review,hasselkamp1992book,Varga1992}
		The discontinuities are due to the discontinuous Galerkin representation of the solution.
		Note that sign has been flipped for the anode velocity coordinate; based on the geometry, the anode distribution is left going and thus has negative velocities.
		}
	\label{f:intro-ion-dist}
\end{figure}

At present, Z-pinch experiments operate at temperatures of approximately \qtyrange{1}{3}{\kilo\electronvolt}\cite{zhang2019sustained,levitt2023,thompson2024} and are expected to scale to tens of \si{\kilo\electronvolt} to achieve higher energy gains.\cite{shumlak2020z}
In addition, the ions are further accelerated toward the wall due to the sheath potential. 
Fig.~\ref{f:intro-ion-dist} shows the ion distribution function for varying bias potentials at the anode and cathode using continuum kinetic simulation data from Ref.~\onlinecite{skolar2023}.
The ions impact the anode at energies of about \SI{0.8}{\kilo\electronvolt} for all bias potentials. 
The cathode is bombarded by significantly higher energy ions as bias potential increases reaching about \SI{11}{\kilo\electronvolt} at a \SI{10}{\kilo\volt} bias potential.
Therefore, IIEE is expected to play a major role in the sheath formation near both Z-pinch electrodes, but especially at the cathode.

It is unclear if the IIEE will be sufficiently strong to shift the configuration from a classical sheath.
Without emissions, a classical sheath forms with a monotonic potential profile, a plasma potential greater than the wall potential, and a positive space charge near the wall.\cite{stangeby2000plasma}
When emissions are sufficiently strong, a space-charge limited (SCL) sheath may form that has a non-monotonic potential profile with an initial decrease below the wall potential before increasing to a plasma potential greater than the wall potential;  
this results in a negative space charge directly near the wall and a positive space charge further from the wall before quasineutrality is reached outside of the sheath region.\cite{hobbs1967SCL,schwager1993SCL,mcadams2012}
If there are sufficient emissions and a cold ion population near the wall, an inverse sheath may form where the monotonic potential profile decreases to a plasma potential lower than the wall potential;
this results in a negative space charge near the wall.\cite{campanell2013negative,campanell2016}

In this work, we build upon previous continuum kinetic sheath simulation results\cite{cagas2017continuum} 
and their associated generalized secondary electron emission boundary condition models.\cite{cagas2020plasma,bradshaw2024,bradshaw2025}
We develop a novel model to obtain the full kinetic ion induced electron emission energy spectrum due to proton, deuteron, or triton bombardment of metallic walls.
We perform 1X-1V (one spatial and one velocity dimension) simulations using the Boltzmann-Poisson system of a doubly bounded plasma between two biased electrodes in present Z-pinch experiment (FuZE)\cite{zhang2019sustained} parameter regimes to expore the effects of IIEE on sheath formation and pinch current.

The paper is organized as follows.
Sec.~\ref{s:ion-impact-model} discusses the IIEE model with comparisons to experimental data.
Sec.~\ref{s:setup} describes the numerical methods and simulation setup.
Sec.~\ref{s:results} presents the results and provides insight into how IIEE changes plasma behavior and Z-pinch currents.
Sec.~\ref{s:discussion} is a discussion on how our results compare with previous emissive sheath literature.
Sec.~\ref{s:conclusions} provides the summary and conclusions of the paper.

\section{Ion Induced Electron Emission (IIEE) Model}  
\label{s:ion-impact-model}

An ion impacting a solid material has the possibility of inducing electron emission.
There are two types of ion induced electron emission (IIEE): potential and kinetic.  
For potential emission, the potential energy released when the impacting ion recombines into a neutral state can result in electron emissions through a variety of processes.\cite{baragiolaPotential,Varga1992} 
For kinetic emission, the kinetic energy of the impacting ion excites an electron in the wall resulting in electron emission.
For velocities greater than \SI[exponent-product=\times,output-exponent-marker=,per-mode=symbol]{2e5}{\meter\per\second}, which is equivalent to proton energies of \SI{209}{\electronvolt}, kinetic emission quickly becomes the dominant IIEE mechanism.\cite{benazeth1982review,hasselkamp1992book,Varga1992}
Fig.~\ref{f:intro-ion-dist} shows that the Z-pinch ion distributions at both electrodes are at velocities higher than the potential-kinetic IIEE boundary (denoted by the vertical dashed blue line).
Thus, we only develop and implement a kinetic emission model.
Ref.~\onlinecite{hasselkamp1992book} provides a fantastic review of the theoretical models and experimental methods for kinetic IIEE.

This section is intended to provide the reader with an understanding of the IIEE model, its validation with experiment, and how to implement as a continuum kinetic boundary condition.
It is organized in the following manner.
Sec.~\ref{s:sey} describes the broader IIEE model to obtain the emitted electron energy spectrum.\cite{schou1980,schou1988}
We use updated models for the low energy ion,\cite{andersen1977} and especially, electron stopping powers\cite{nguyentruong2015}, which are described in Secs.~\ref{s:ion-dEdx} and \ref{s:elc-dEdx}, respectively.
Sec.~\ref{s:gamma_m} explains how we calculate the scaling parameter.\cite{schou1980}
Sec.~\ref{s:spectrum-validation} examines how well the improved model for the emitted electron energy spectrum matches data.
In Sec.~\ref{s:gaussian}, we explain how to fit the emission spectrum to a Gaussian curve to calculate the SEY.
Sec.~\ref{s:SEY_validation} validates the model against experimental SEY data.
Lastly, Sec.~\ref{s:elc-dist} explains how to obtain the emitted electron distribution function from the emitted electron energy spectrum for use in continuum kinetic simulations.\cite{bradshaw2024}\textit{}

\subsection{Secondary Electron Yield and the Emitted Electron Energy Spectrum} 
\label{s:sey}
Secondary electron emission can be characterized through the secondary electron yield (SEY), which is
\begin{equation}
	\gamma = \frac{\Gamma_{e,emit}}{\Gamma_{i,impact}},
	\label{eq:gamma}
\end{equation}
where $\Gamma_{e,emit}$ is the emitted electron particle flux
and $\Gamma_{i,impact}$ is the impacting species particle flux, which in our case are ions.

Several different models have been developed for kinetic electron emissions due to ion impact.
The early work by Ref.~\onlinecite{sternglass1957} found that the SEY is proportional to the ion stopping power. 
However, this model is only valid for proton impact above approximately \SI{100}{\kilo\electronvolt}\cite{hasselkamp1992book} due to a limited understanding of low energy ion and electron stopping powers at the time.  
Other models have extended this work to lower energies.\cite{beuhler1977}
Semi-empirical models have also been developed to obtain the SEY.\cite{baragiola1979,stoltz2003,haque2019}

While the SEY is useful to understand, it is a bulk property and cannot describe the kinetic effects of electron emission.
For this, we need the emitted electron energy spectrum, $d\gamma/dE_e$, where $E_e$ is the emitted electron energy.
While other semi-empirical and theoretical models exist for obtaining the emitted electron energy spectrum due to IIEE\cite{hasselkamp1992book},  
we adapt the ionization cascade model from Ref.~\onlinecite{schou1980}.
This work is an improvement over previous models due to its closed form expressions, treatment of the electron transport through a cascade process, and increased energy range of validity.\cite{hasselkamp1992book} 
Due to the requirement of a sufficiently large number of energetic electrons and a limited understanding of the low energy electron stopping power at the time, the original model\cite{schou1980} is valid above ion impacting energies of several \si{\kilo\electronvolt}.\cite{schou1988}
This threshold is generally in line with the expected ion energies as per Fig.~\ref{f:intro-ion-dist}(b).
In addition, significant improvements have been made to the understanding of the low energy electron stopping power\cite{nguyentruong2015} (discussed in more detail in Section~\ref{s:elc-dEdx}), which should improve the lower energy range of validity for the IIEE model.

The ionization cascade mechanism assumes a free electron gas within the wall limiting the choice of materials to primarily metals.
The steps of the mechanism are as follows. 
An energetic ion impacts and enters the wall material.
As it passes through, it may collide with an atom or molecule of the wall, possibly exciting an electron.
The excited electron is ejected in a random direction and begins traveling through the wall.
This excited electron may then collide with another wall atom or molecule, which may excite yet another electron which is then ejected in a random direction.
This process continues resulting in a cascading effect.
The total spectrum of electrons that exit the wall in the direction of the origin of the impacting ion is\cite{schou1980}
\begin{equation}
	\frac{d\gamma}{dE_e} = \underbrace{S_i(E_i)}_{\text{projectile}} 
	\underbrace{\frac{\Gamma_r E_e}{4 [E_e + W]^2 S_e(E_e+W) }}_{\text{wall}} ,
	\label{eq:dgdE}
\end{equation}
where $S_i(E_i)$ is the ion stopping power for an ion with energy $E_i$,
$\Gamma_r$ is a scaling parameter,
$E_e$ is the emitted electron energy,
$W$ is the barrier height defined as the summation of the work function ($\phi_w$) and the Fermi energy ($E_F$),
and $S_e(E_e+W)$ is the electron stopping power evaluated for an electron with energy $E_e+W$.
The work function and Fermi energy can be found in standard material tabulations.\cite{crc2018}
The methods of obtaining the remaining parameters are discussed in Secs.~\ref{s:ion-dEdx}, \ref{s:elc-dEdx}, and \ref{s:gamma_m}.
For this work, only the electronic (or inelastic) stopping power is considered for both the ions and electrons because it is expected to be larger than the nuclear (or elastic) stopping power in these energy regimes.
For all future uses of stopping power, assume that it is the electronic stopping power.

A utility of Eq.~\ref{eq:dgdE} is that it can be split into two parts: the projectile and wall parts.
Thus, from a computational perspective, the wall part can be entirely pre-computed and only the projectile part needs to be considered at each time step.
Furthermore, the shape of the emitted electron energy spectrum is dependent entirely on the wall component with the projectile component only acting as a scaling parameter. This is a useful property that is exploited in several ways, as will be discussed later in Sec.~\ref{s:gaussian} and \ref{s:elc-dist}.

\subsection{Ion Stopping Power}
\label{s:ion-dEdx}
Traditionally, the ion stopping power is calculated using the Bethe formula\cite{bethe1930} with a myriad of corrections.\cite{icru49_2,bloch1933,barkas1963,lindhard1976}  
However, these formulations are only accurate at high energies with a lower limit in the regime of hundreds of \si{\kilo\electronvolt}, which is much larger than the expected Z-pinch ion energies.
For low energy protons, deuterons, or tritons, an empirical model\cite{andersen1977} exists for obtaining the ion stopping cross sections, which is defined as the stopping power divided by the wall material number density.
The cutoff energies below which this empirical model is valid are shown in Table 3.7 from Ref.~\onlinecite{icru49_3}.
Between all the materials listed, the lowest cutoff energy is at \SI{100}{\kilo\electronvolt}; furthermore, fusion electrode relevant materials such as graphite and copper have cutoff energies of \SI{200}{\kilo\electronvolt} and \SI{500}{\kilo\electronvolt}, respectively. 
Since all of the ion energies in the Z-pinch plasma are predicted to be less than \SI{100}{\kilo\electronvolt} per Fig.~\ref{f:intro-ion-dist} and Z-pinch scaling studies\cite{shumlak2020z}, this empirical model\cite{andersen1977} is used for all ion stopping power calculations.  
Converted to ion stopping power in SI units of \si[per-mode=symbol]{\joule\per\meter}, the formulation is
\begin{equation}   
	S_i = \frac{\epsilon_{\text{low}} \epsilon_{\text{high}}}{
		\epsilon_{\text{low}} + \epsilon_{\text{high}}}  \frac{n_we}{10^{19}},
	\label{eq:low-energy-ion-stopping-power}
\end{equation}
where the $n_w$ is the wall material number density in \si{\meter^{-3}}, 
$e$ is the elementary electric charge in \si{\coulomb},
and $\epsilon_{\text{low},\text{high}}$ are empirical formulas
of the ion stopping cross section
for varying regimes of ion energies.\cite{andersen1977,icru49_3}
The wall material number density is defined as $n_w=\rho_w/m_wu$, where
$\rho_w$ is the wall material mass density in \si[per-mode=symbol]{\kilogram\per\meter^3}, 
$m_w$ is the wall material atomic mass in \si{amu},
and $u$ is the unified atomic mass unit.
The empirical formulas for the low and high ion stopping cross sections are\cite{andersen1977}
\begin{eqnarray}  
	\epsilon_{\text{low}} &&= A_2 T_s^{0.45} 
	\label{eq:epsilon-low} \\
	\epsilon_{\text{high}} &&= \frac{A_3}{T_s} 
	\ln \bigg[ 1 + \frac{A_4}{T_s} + A_5 T_s  \bigg],
	\label{eq:epsilon-high}
\end{eqnarray} 
where $A_{2-5}$ are empirically determined coefficients that are found in Table 3.1 from Ref.~\onlinecite{icru49_3} 
and $T_s=E_iu/m_{p,d,t}$ is the scaled ion energy where $E_i$ is the ion energy in \si{\kilo\electronvolt}, 
$u$ is the unified atomic mass unit, and $m_{p,d,t}$ is the mass of a proton, deuteron, or triton in \si{\kilogram}.
This model is valid down to ion energies of \SI{1}{\kilo\electronvolt} with an accuracy of 20-30\%. 
Despite the inaccuracy, it is found that using this empirical ion stopping power model accurately captures the SEY for more well studied materials such as aluminum, gold, and copper.
This is further discussed in Sec.~\ref{s:SEY_validation}.

This stopping power model is only valid for proton, deuteron, and triton impacts and has utility in its analytical form and simplicity.
Other ion species will require different models to properly calculate their low energy stopping powers.
For example, the Monte Carlo collision code TRIM (Transport of Ions in Matter)\cite{srim2010}
can calculate the average nuclear and electronic stopping powers for arbitrary ion species and wall materials.

\subsection{Electron Stopping Power}
\label{s:elc-dEdx}
The Bethe formula\cite{bethe1930,bloch1933,barkas1963,icru49_2,lindhard1976} is highly inaccurate at low electron energies and predicts negatively valued stopping powers.\cite{nguyentruong2015} 
Despite Ref.~\onlinecite{schou1980} providing a simple formulation for the low energy electron stopping power for use in the IIEE model, it is not very accurate.
Significant improvements have since been made with an empirical modification to the original Bethe formula that is generalizable based on material constants, always positive, and better fits the data.\cite{nguyentruong2015}
In the range of 50-\SI{600}{\electronvolt}, the electron stopping power model has an error of 10-20\% depending on the material with copper having an error of 20.60\%.
Extending the range to \SI{30}{\kilo\electronvolt} improves overall accuracy with copper's error going to 9.96\%.
Furthermore, due to the properties of this model, it is able to make predictions of the electron stopping power at arbitrarily low energies.
Based on experimental data,\cite{hasselkamp1983spectrum} the emitted electron energies for copper are generally less than \SI{1}{\kilo\electronvolt}.
The modified Bethe formula for low energy electron stopping power,\cite{nguyentruong2015} in SI units of \si[per-mode=symbol]{\joule\per\meter}, is
\begin{equation}
	S_e = \frac{e^4 Z n_w}{ 8 \pi \epsilon_0^2 E} 
	\ln \bigg[  \sqrt{\frac{\mathtt{e}}{2}} \frac{E}{I}  + G(E)\bigg], 
	\label{eq:bethe_nguyen_correction_SI}
\end{equation}
where $e$ is the elementary electric charge in \si{\coulomb},
$Z$ is the wall material atomic number,
$n_w$ is the wall material number density in \si{\meter^{-3}},
$\epsilon_0$ is the vacuum permittivity,
$E$ is the electron energy in \si{\joule},  
$\mathtt{e}$ is the base of the natural logarithm,  
$I$ is the mean excitation energy in \si{\joule}, and
\begin{widetext}  
	\begin{equation}
		G(E) = 1 - \sqrt{\frac{\mathtt{e}}{2}} \ln \bigg[ 1 + \Big( \frac{E}{I} \Big)^2 \bigg] \frac{I}{E} 
		+ \frac{1}{3} \ln \Big( \frac{Z}{2} \Big) 
		\exp \bigg[ - \frac{3}{\sqrt{Z}} \Big( 1 - \frac{2}{\sqrt{Z}} + \ln \frac{E}{I} \Big)^2  \bigg] \frac{E}{I}.
		\label{eq:G_nguyen}
	\end{equation}
\end{widetext}
It is important to note that when the electron stopping power is evaluated in the IIEE model (Eq.~\ref{eq:dgdE}), the energy used should be the summation of the electron energy and the barrier height: $E=E_e+W$.

\subsection{Scaling Parameter}
\label{s:gamma_m}
With the ion and electron stopping powers, Eq.~\ref{eq:dgdE} can be calculated up to the constant $\Gamma_r$, which is defined as\cite{schou1980}
\begin{equation}
	\Gamma_r = \frac{r}{\psi(1) - \psi(1-r) }. \label{eq:Gamma_m}
\end{equation}
where $\psi(x) = d(\ln [\Gamma (x)])/dx$ is the zeroth order polygamma function,
$\Gamma(x)$ is the gamma function, 
and $r$ is related to the barrier height and the energy of peak emission, $E_{peak}$, by\cite{schou1980}
\begin{equation}
	r = 2 - \frac{1}{2} \frac{W}{E_{peak}}.
	\label{eq:m}
\end{equation}
The energy of peak emission can be found by computing the wall term of Eq.~\ref{eq:dgdE} without $\Gamma_r$ and finding the energy at which the maximum value occurs. 

\subsection{Emission Spectrum Model Validation}
\label{s:spectrum-validation}

The wall material parameters required to calculate Eq.~\ref{eq:dgdE} are the
atomic number ($Z$), 
number density ($n_w$),
work function ($W$),
and mean excitation energy ($I$).
These material parameters are tabulated for a variety of elemental metals in Table~\ref{tab:material-param}.
For our calculations, we use the average of all listed work functions for each material.\cite{crc2018}
The mean excitation energies are obtained from Ref.~\onlinecite{seltzer1982}.
In addition, Table~\ref{tab:material-param} lists the empirical ion stopping power coefficients\cite{icru49_3} used in Eqs.~\ref{eq:low-energy-ion-stopping-power}-\ref{eq:epsilon-high}.
The work function and Fermi energy for tungsten, which is a fusion relevant wall material, are obtained from Refs.~\onlinecite{michaelson1977} and \onlinecite{lemell2009}, respectively.

\begin{table*}[!htb]
	\caption{\label{tab:material-param}
		A table with material parameters\cite{crc2018,seltzer1982,michaelson1977,lemell2009},
		empirical ion stopping power coefficients\cite{icru49_3},
		and Gaussian fitting parameters
		for various elemental metals.
	}
	\begin{ruledtabular}
		\begin{tabular}{c|cccc|cccc|ccc}
			$\ $ & \multicolumn{4}{c|}{Material Parameters\cite{crc2018,seltzer1982,michaelson1977,lemell2009}} & \multicolumn{4}{c|}{Ion Stopping Power Coefficients\cite{icru49_3}} & \multicolumn{3}{c}{Gaussian Fitting Parameters} \\ \hline 
			\rule{0pt}{10pt}\vspace{1pt} Element & $Z$ & $n_w$ (\si[per-mode=symbol]{\kilogram\per\meter^3}) & $W$ (\si{\electronvolt}) & $I$ (\si{\electronvolt}) &   $A_2$ & $A_3$ & $A_4$ & $A_5$ & $E_0$ (\si{\electronvolt}) & $\tau$ & $\int_0^\infty g(E_e) dE_e$ (\si[per-mode=symbol]{\meter\per\joule})
			\\ 
		 	\hline
			Li  & 3 & \num{4.633e28} & 7.670 & 40  & 1.600 & \num{7.256e2} & 3013  & \num{4.578e-2} & 2.056  & 1.233 & \num{1.147e8} \\
			Na & 11 & \num{2.536e28} & 5.6   & 149 & 2.869 & \num{2.628e3} & 1854  & \num{1.472e-2} & 1.201  & 1.014 & \num{2.835e9} \\
			Mg & 12 & \num{4.306e28} & 10.74 & 156 & 4.293 & \num{2.862e3} & 1009  & \num{1.397e-2} & 2.243  & 1.167 & \num{2.180e8} \\
			Al & 13 & \num{6.026e28} & 15.87 & 166 & 4.739 & \num{2.766e3} & 164.5 & \num{2.023e-2} & 3.008  & 1.562 & \num{7.440e7} \\
			K  & 19 & \num{1.371e28} & 4.41  & 190 & 5.833 & \num{4.482e3} & 545.7 & \num{1.129e-2} & 0.8611 & 1.046 & \num{4.659e9} \\
			Ca & 20 & \num{2.329e28} & 7.56  & 191 & 6.252 & \num{4.710e3} & 553.3 & \num{1.112e-2} & 1.619  & 1.139 & \num{4.020e8} \\
			Mn & 25 & \num{7.903e28} & 14.56 & 272 & 3.907 & \num{5.725e3} & 1461  & \num{8.829e-3} & 3.450  & 1.352 & \num{4.044e7} \\
			Fe & 26 & \num{8.476e28} & 15.84 & 286 & 3.963 & \num{6.065e3} & 1243  & \num{7.782e-3} & 3.727  & 1.413 & \num{3.443e7} \\
			Cu & 29 & \num{8.491e28} & 11.76 & 322 & 4.194 & \num{4.649e3} & 81.13 & \num{2.242e-2} & 2.807  & 1.210 & \num{6.337e7} \\  
			Zn & 30 & \num{6.576e28} & 13.74 & 330 & 4.750 & \num{6.953e3} & 295.2 & \num{6.809e-3} & 3.377  & 1.275 & \num{5.684e7} \\
			Ga & 31 & \num{5.105e28} & 14.72 & 334 & 5.697 & \num{7.173e3} & 202.6 & \num{6.725e-3} & 3.656  & 1.322 & \num{6.023e7} \\
			Rb & 37 & \num{1.079e28} & 4.111 & 363 & 6.429 & \num{8.478e3} & 292.9 & \num{6.087e-3} & 0.8357 & 1.070 & \num{6.659e9} \\
			Sr & 38 & \num{1.814e28} & 6.52  & 366 & 7.159 & \num{8.693e3} & 330.3 & \num{6.003e-3} & 1.467  & 1.120 & \num{8.709e8} \\
			Nb & 41 & \num{5.555e28} & 9.649 & 417 & 7.791 & \num{9.333e3} & 442.7 & \num{5.587e-3} & 2.368  & 1.176 & \num{1.192e8} \\
			Ag & 47 & \num{5.856e28} & 10.12 & 470 & 6.038 & \num{6.790e3} & 397.8 & \num{1.676e-2} & 2.573  & 1.189 & \num{9.521e7} \\
			Cd & 48 & \num{4.634e28} & 11.55 & 469 & 6.554 & \num{1.080e4} & 355.5 & \num{4.626e-3} & 3.035  & 1.226 & \num{8.250e7} \\
			In & 49 & \num{3.834e28} & 12.72 & 488 & 7.024 & \num{1.101e4} & 370.9 & \num{4.540e-3} & 3.397  & 1.248 & \num{8.573e7} \\
			Sn & 50 & \num{3.686e28} & 14.62 & 488 & 7.227 & \num{1.121e4} & 386.4 & \num{4.474e-3} & 3.992  & 1.308 & \num{6.412e7} \\
			Sb & 51 & \num{3.312e28} & 15.53 & 487 & 8.480 & \num{8.608e3} & 348.0 & \num{9.074e-3} & 4.267  & 1.348 & \num{6.056e7} \\
			Cs & 55 & \num{8.745e27} & 3.54  & 488 & 8.218 & \num{1.223e4} & 399.7 & \num{4.447e-3} & 0.7662 & 1.083 & \num{7.237e9} \\
			Ba & 56 & \num{1.539e28} & 6.16  & 491 & 8.911 & \num{1.243e4} & 402.1 & \num{4.511e-3} & 1.495  & 1.136 & \num{7.809e8} \\
			W  & 74 & \num{6.322e28} & 15.15 & 727 & 5.160 & \num{1.540e4} & 415.0 & \num{3.410e-3} & 4.440  & 1.292 & \num{3.468e7} \\
			Au & 79 & \num{5.901e28} & 10.91 & 790 & 5.458 & \num{7.852e3} & 975.8 & \num{2.077e-2} & 3.065  & 1.209 & \num{7.431e7} \\
			Hg & 80 & \num{4.063e28} & 11.61 & 800 & 4.843 & \num{1.704e4} & 487.8 & \num{2.882e-3} & 3.302  & 1.220 & \num{9.484e7} \\
			Tl & 81 & \num{3.492e28} & 11.99 & 810 & 5.311 & \num{1.722e4} & 537.0 & \num{2.913e-3} & 3.437  & 1.227 & \num{1.025e8} \\
			Pb & 82 & \num{3.296e28} & 13.72 & 823 & 5.982 & \num{1.740e4} & 586.3 & \num{2.871e-3} & 4.024  & 1.255 & \num{8.415e7} \\
			Bi & 83 & \num{2.818e28} & 14.24 & 823 & 6.700 & \num{1.780e4} & 677.0 & \num{2.660e-3} & 4.215  & 1.268 & \num{8.807e7} \\
		\end{tabular}
	\end{ruledtabular}
\end{table*}

\begin{figure}[!htb] 
	\centering
	\includegraphics[width=\linewidth]{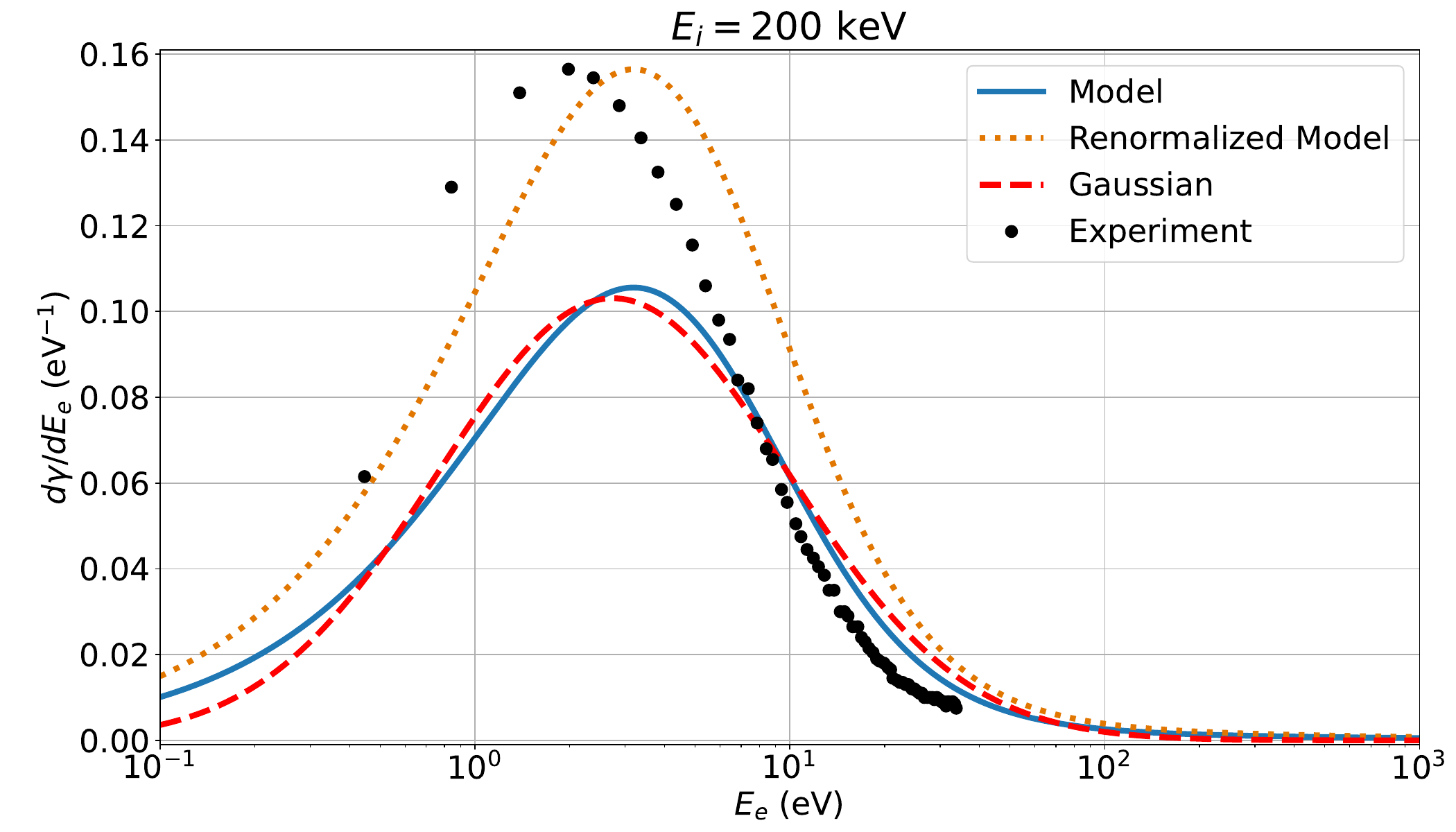}
	\caption{
		Comparison of emitted electron energy spectrum from the model (blue solid line), renormalized model (orange dotted line), Gaussian fit (red dashed line), and data\cite{hasselkamp1983spectrum} (black circles) for an incoming monoenergetic \SI{200}{\kilo\electronvolt} proton beam striking a copper target.
	}
	\label{f:spectrum-data-gauss-compare}
\end{figure}

Therefore, using Eqs.~\ref{eq:low-energy-ion-stopping-power}-\ref{eq:m} and material parameters from Table~\ref{tab:material-param}, the emitted electron energy spectrum, Eq.~\ref{eq:dgdE}, can be calculated. 
The blue line in Fig.~\ref{f:spectrum-data-gauss-compare} shows an example plot of the emitted electron energy spectrum caused by a monoenergetic \SI{200}{\kilo\electronvolt} proton beam striking a copper target. 
The model provides a significant underprediction of the emission spectrum compared to the data\cite{hasselkamp1983spectrum} (black circles).
As such, this model has traditionally been used by renormalizing to the data\cite{schou1980,schou1988} (dotted orange line).
The renormalized model has a higher peak energy of \SI{3.188}{\electronvolt} compared to that of the data which is \SI{1.980}{\electronvolt}.
This matches expected behavior based on Fig. 5 from Ref.~\onlinecite{schou1988} and comparisons to energy of peak emission data from Ref.~\onlinecite{hasselkamp1986peak}.
Despite the inaccuracies, the model is useful due to a general lack of experimental data for emitted electron energy spectra from ion impacts.  
This model provides a general approximation of the emitted spectrum without the need for an extensive dataset.

\begin{figure}[!htb] 
	\centering
	\includegraphics[width=\linewidth]{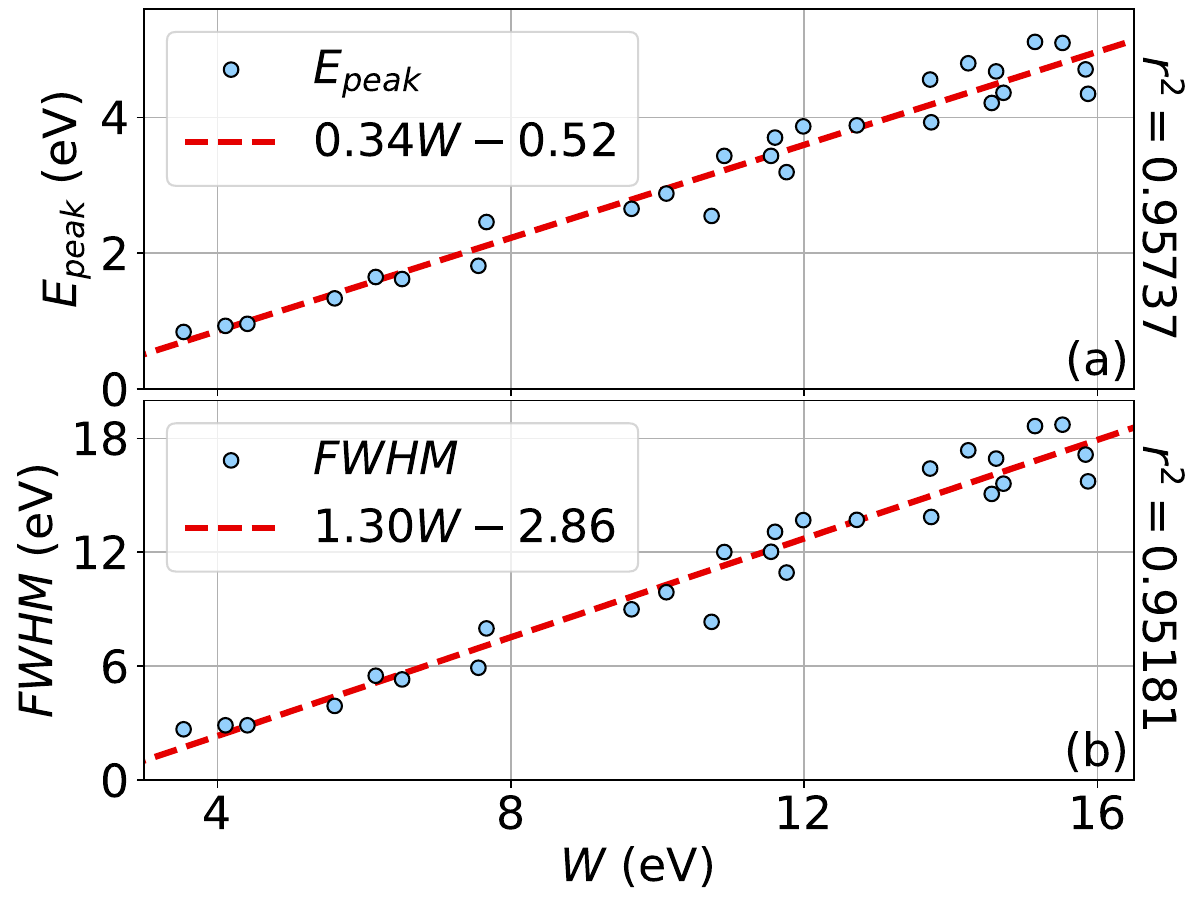}
	\caption{
		The energy of peak emission (a) and full width half max (b) calculated based on the wall portion of Eq.~\ref{eq:dgdE} for all materials listed in Table~\ref{tab:material-param}. 
		Shown as a function of the barrier height.
		Both panels show a positive linear trend.
	}
	\label{f:model-trends-with-W}
\end{figure}

In addition, several key features of the model can be generally described by the barrier height.
Fig.~\ref{f:model-trends-with-W} shows the energy of peak emission and the full width half max ($FWHM$) of the wall portion of Eq.~\ref{eq:dgdE} as a function of the barrier height for all of the materials listed in Table~\ref{tab:material-param}.
Both values show a strong positive linear correlation with barrier height.
For all of these materials, the energy of peak emission is in the range of \qtyrange{1}{6}{\electronvolt}. 
Note that this energy is substantially smaller than the thermal electron energy in a Z-pinch plasma, which is on the order of \qtyrange{1}{3}{\kilo\electronvolt}.\cite{levitt2023,thompson2024}

\subsection{Calculating SEY Using Gaussian Fits}
\label{s:gaussian}

While there are few IIEE emitted electron energy spectra datasets, a plethora of data exists for the secondary electron yield (SEY)\cite{hasselkamp1992book,haque2019} with which to better validate the model.
The SEY (Eq.~\ref{eq:gamma}) is found by integrating Eq.~\ref{eq:dgdE} with respect to electron energy from 0 to $\infty$, 
$\gamma=\int_0^\infty (d\gamma/dE_e) dE_e$.
Only the wall portion of Eq.~\ref{eq:dgdE} is a function of the electron energy resulting in a simplified integral of
\begin{equation}
	\gamma = S_i(E_i) \underbrace{\int_0^\infty  \frac{\Gamma_r E_e}{4 [E_e + W]^2 S_e(E_e+W) } dE_e}_{\text{constant}}.
	\label{eq:gamma-integral}
\end{equation}
Because every term in the integrand is either a wall material parameter or dependent solely on electron energy, the integral in Eq.~\ref{eq:gamma-integral} becomes a constant.
This is in line with previous semi-empirical and theoretical works showing that the SEY is proportional to the ion stopping power.\cite{sternglass1957,stoltz2003,haque2019}

The integral is complicated and lacks an analytical solution.
Therefore, it is evaluated numerically using the trapezoid rule.
Interestingly, despite the energy spectrum visually appearing to go to zero as the energy increases, as shown in Fig.~\ref{f:spectrum-data-gauss-compare}, the integral does not converge as the numerical upper bound is taken to $\infty$.
This is either because $d\gamma/dE_e$ tends to 0 very slowly as $E_e$ goes to $\infty$, or it tends to a small but finite number as $E_e$ goes to $\infty$.   
Regardless, this has the effect of the integral tending to $\infty$.
Realistically, the SEY cannot go to $\infty$.
To resolve this, we fit the wall part of the emitted electron energy spectrum (Eq.~\ref{eq:dgdE}) to a Gaussian of a logarithm (to be just called a Gaussian from henceforth) of the form\cite{scholtz1996,bradshaw2024}
\begin{equation}  
	g(E_e) = C \exp \left( - \displaystyle\frac{ \bigg[ \ln \Big( \frac{E_e}{E_0} \Big) \bigg]^2 }{2 \tau^2} \right),
	\label{eq:gauss}
\end{equation}
where $C$, $E_0$, and $\tau$ are fitting parameters that correspond to the scale, location in energy, and width of the emitted electron energy spectrum, respectively.
Thus, the emitted electron energy spectrum becomes 
\begin{equation}
	\frac{d\gamma}{dE_e} = \underbrace{S_i(E_i)}_{\text{projectile}} \underbrace{g(E_e)}_{\text{wall}}.
	\label{eq:dgdE_gauss}
\end{equation}
The red dashed line in Fig.~\ref{f:spectrum-data-gauss-compare} shows that Eq.~\ref{eq:dgdE_gauss} is a good approximation of Eq.~\ref{eq:dgdE} (solid blue line). 

Furthermore, Eq.~\ref{eq:gauss} has a finite integral from 0 to $\infty$:
\begin{equation}
	\int_0^\infty g(E_e) dE_e = C E_0 \tau (2\pi)^\frac{1}{2} \exp \bigg( \frac{\tau^2}{2} \bigg).
	\label{eq:gauss-integral}
\end{equation}
Substituting Eq.~\ref{eq:gauss-integral} into Eq.~\ref{eq:gamma-integral} results in an SEY of
\begin{equation}
	\gamma =  C E_0 \tau (2\pi)^\frac{1}{2}\exp \bigg( \frac{\tau^2}{2} \bigg) S_i(E_i).
	\label{eq:gamma_gauss}
\end{equation}
Therefore, the SEY is proportional to the ion stopping power scaled by the Gaussian fitting coefficients that depend entirely on the wall material parameters.

The fitting coefficients are found through the nonlinear least squares method using \verb|scipy.optimize.curve_fit()| in Python.
From a practical perspective, it is easier to fit to a normalized version of Eq.~\ref{eq:dgdE} that is divided by its maximum value such that the range is only from 0 to 1. 
The fit is done over the electron energy range of \qtyrange{0}{1500}{\electronvolt} with \num[exponent-product=\times,output-exponent-marker=]{e5} data points.
The resulting multiplicative fitting coefficient can then be re-scaled to obtain $C$.
Eq.~\ref{eq:gauss-integral} is calculated using the resulting fitting coefficients.

Table~\ref{tab:material-param} also provides the Gaussian fitting parameters ($E_0$ and $\tau$) and the integral of the Gaussian (Eq.~\ref{eq:gauss-integral}) for all materials.
If needed, the coefficient $C$ can be found by solving Eq.~\ref{eq:gauss-integral} for $C$ and using the known table values.

\subsection{SEY Model Validation}
\label{s:SEY_validation}

\begin{figure*}[!htb]
	\centering
	\includegraphics[width=\linewidth]{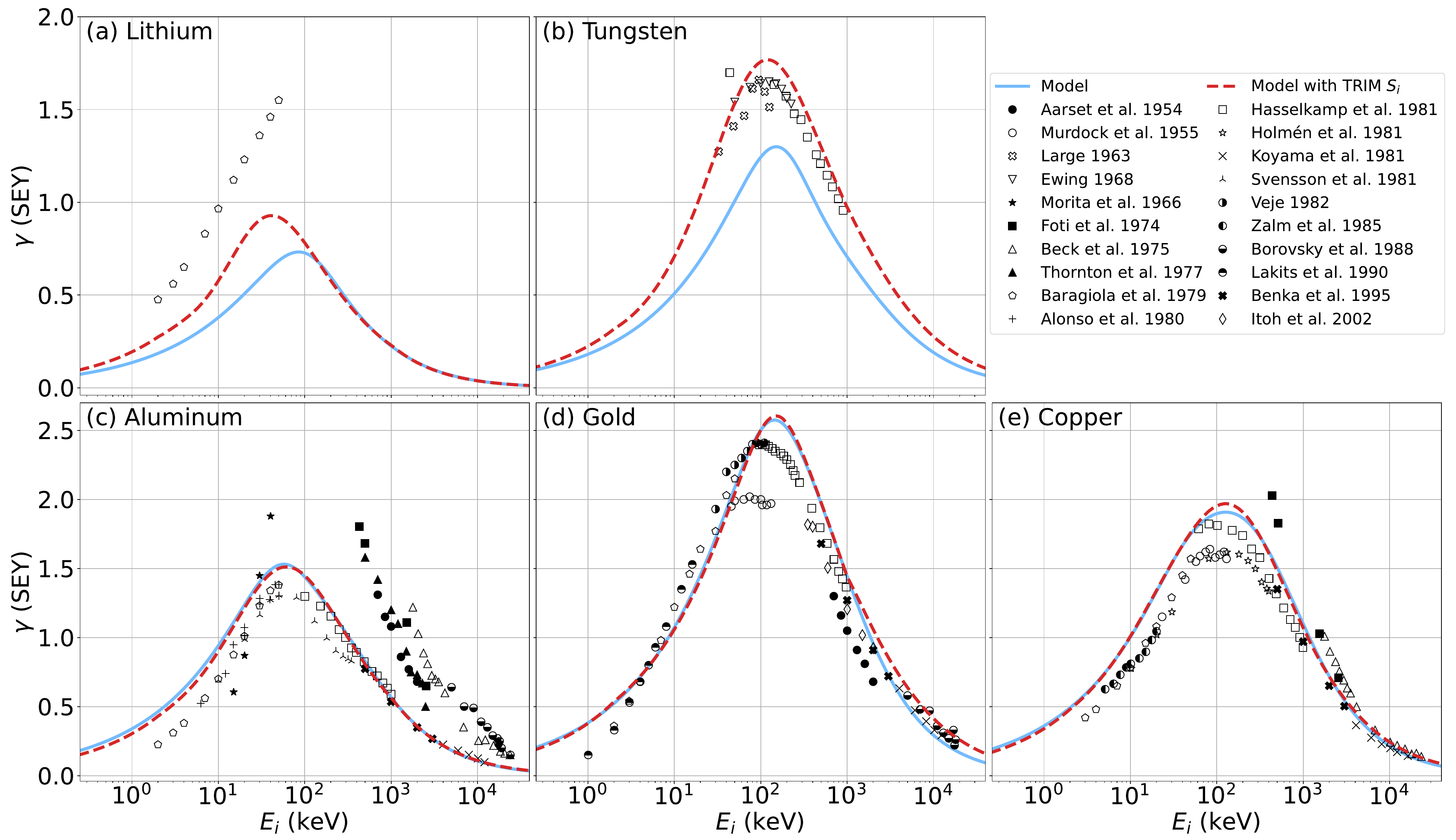}
	\caption{Comparison of model SEY (blue line) to experimental data (markers) for 
		lithium,\cite{baragiola1979}
		tungsten,\cite{large1963,ewing1968,hasselkamp1981}
		aluminum,\cite{aarset1954,morita1966,foti1974,beck1975,thornton1977,baragiola1979,alonso1980,hasselkamp1981,koyama1981,svensson1981,borovsky1988,benka1995} 
		gold,\cite{aarset1954,murdock1955,baragiola1979,hasselkamp1981,koyama1981,veje1982,borovsky1988,lakits1990,benka1995,itoh2002}
		and copper,\cite{murdock1955,foti1974,beck1975,baragiola1979,baragiola1979,hasselkamp1981,holmen1981,koyama1981,zalm1985,benka1995} 
		The dashed red line corresponds to a similar calculation but with ion stopping powers calculated using TRIM.\cite{srim2010}
	}
	\label{f:model-comparison-gamma}
\end{figure*}

The IIEE model can be further validated by comparing the SEY calculated using Eq.~\ref{eq:gamma_gauss} to experimental data.
Fig.~\ref{f:model-comparison-gamma} shows the comparison of the model SEY (blue line) to experimental data.\cite{aarset1954,murdock1955,large1963,ewing1968,morita1966,foti1974,beck1975,thornton1977,baragiola1979,alonso1980,hasselkamp1981,holmen1981,koyama1981,svensson1981,veje1982,zalm1985,borovsky1988,lakits1990,benka1995,itoh2002}
For the well-studied metals (aluminum, copper, and gold), the model generally agrees with the data.
For other materials that are more fusion relevant (lithium and tungsten), the model severely underpredicts the SEY compared to the data.

The difference between the model and the (notably sparse) data for lithium may be due to lithium's tendency to oxidize quickly.
``Dirty" or oxidized materials have higher IIEE SEY\cite{hasselkamp1986peak,bogaerts2002}.
This phenomenon has also been shown for electron induced secondary electron emission in lithium.\cite{capece2016,bradshaw2025}
In addition, Ref.~\onlinecite{hasselkamp1992book} notes that several of the datasets shown in Fig.~\ref{f:model-comparison-gamma} used contaminated samples.
Note how some of the data points on the right portion of Fig.~\ref{f:model-comparison-gamma}(c) appear to be following a slightly higher curve than the other points;
this is because they are likely taken from an oxidized sample.
Borovsky et al. (1988)\cite{borovsky1988} (bottom-filled circles) actually provide data for aluminum oxide instead of aluminum and follow the same upper curve.

Since the SEY is proportional to the ion stopping powers, our model can be improved with better ion stopping power calculations.
For example, a calculation is provided using TRIM\cite{srim2010} (red dashed line in Fig.~\ref{f:model-comparison-gamma}) with good agreement to our model and the data for aluminum, gold, and copper.
In addition, a notable improvement is made for tungsten and better matches experiment.
While still severely underpredicting the SEY for lithium, the TRIM results are closer to the data than our model is.
Other low energy ion stopping power models\cite{haque2019} may yield further improved SEY.

For comparisons to other materials, Refs.~\onlinecite{hasselkamp1992book,haque2019} provide extensive sources for experimental data.

\subsection{Emitted Electron Distribution Function}
\label{s:elc-dist}
The Gaussian fit (Eq.~\ref{eq:gauss}) provides the emitted electron energy spectrum. 
However, for evolving a plasma kinetically with the Vlasov or Boltzmann equation, we need the emitted electron distribution function.
The relationship between the emitted electron energy spectrum and the emitted electron energy distribution function is found by taking the derivative of Eq.~\ref{eq:gamma} with respect to the electron energy:
\begin{equation}
	\frac{d\gamma}{dE_e} = 
	\frac{d}{dE_e} \bigg(\frac{\Gamma_{e,emit}}{\Gamma_{i,impact}} \bigg).
	\label{eq:dgdE_gamma}
\end{equation}
The impacting ion particle flux is not a function of electron energy and can be taken outside of the derivative.
The electron energy can be related to the velocity through the kinetic energy as $E_e = m_e v_e^2/2$.
After using chain rule to take the derivative with respect to $v_e$ instead of $E_e$, 
using the fact that the electron particle flux is the first moment of the electron distribution function ($\Gamma_e=\int f_e v_e dv_e$), and some algebraic manipulation, the relationship between the emitted electron energy spectrum and the emitted electron distribution function, $f_{e,emit}$, is\cite{bradshaw2024}
\begin{equation}
	f_{e,emit} = m_e \Gamma_{i,impact} \frac{d\gamma}{dE_e}.
	\label{eq:rel-fe-dgammadE}
\end{equation}
Thus, the distribution function is obtained by scaling the energy spectrum by the electron mass and ion particle flux.

With a combination of Eqs.~\ref{eq:dgdE_gauss} and \ref{eq:rel-fe-dgammadE}, the emitted electron distribution function can be obtained.
However, the original model\cite{schou1980} assumes that the impacting ions are due to a monoenergetic ion beam.
In our continuum kinetic simulations, we have a distribution of ion energies instead. 
To account for this difference, the discretized ion distribution function, $f_{i,impact}^j$ (indexed by $j$), can be thought of as a set of individual monoenergetic ion beams with energy $E_i^j$ and velocity $v_i^j$. 
For a uniform velocity mesh, the associated impacting ion particle flux is $\Gamma_{i,impact}^j = v_i^j f_{i,impact}^j \Delta v$, where $\Delta v = v_i^{j+1}-v_i^j$.
The total emitted electron distribution function can be found by taking the summation of each individual emitted electron distribution function (Eq.~\ref{eq:rel-fe-dgammadE}) caused by impacting particle flux $\Gamma_{i,impact}^j$.
Therefore, substituting Eq.~\ref{eq:dgdE_gauss} into Eq.~\ref{eq:rel-fe-dgammadE} and summing over all $j$ yields\cite{bradshaw2024}
\begin{equation}
	f_{e,emit}^{TOT} = m_e g(E_e) \Delta v
			\sum_j v_i^j f_{i,impact}^j S_i(E_i^j).
			\label{eq:feTotal}
\end{equation}

\section{Numerical Methods and Simulation Setup}


\label{s:setup}

The Z-pinch is modeled as a proton-electron plasma bounded between two walls with an electric potential bias between them.
The plasma is evolved kinetically using the 1X-1V Boltzmann equation,
\begin{equation}
	\frac{\partial f_\alpha}{\partial t} + \frac{\partial (f_\alpha v_x)}{\partial x} 
	+ \frac{q_\alpha}{m_\alpha} \frac{\partial (f_\alpha E_x)}{\partial v_x} = 
	\sum \Big( \frac{\partial f_\alpha}{\partial t} \Big)_c + S_\alpha,
	\label{eq:boltzmann}
\end{equation}
where $f$ is the distribution function, $q$ is the electric charge, and $m$ is the mass for species $\alpha$ (ions or electrons).

The summation term is the Dougherty (sometimes called the Lenard-Bernstein) collision operator\cite{dougherty1964model,francisquez2020conservative,hakim2020conservative} and includes the effects of all inter- and intra-species elastic collisions where the collision frequencies are set to
\begin{align}
	\nu_{ee} &= \frac{v_{th_e}}{\lambda_{MFP}} \label{eq:nuee} \\
	\nu_{ei} &= \nu_{ee}\sqrt{2} \label{eq:nuei}\\
	\nu_{ii} &= \frac{v_{th_i}}{\lambda_{MFP} } \label{eq:nuii}\\
	\nu_{ie} &= \frac{m_e}{m_i} \nu_{ee} \label{eq:nuie} ,
\end{align}
where $\lambda_{MFP}$ is the mean free path and $v_{th_\alpha}$ is the thermal velocity (defined as $v_{th_\alpha}=\sqrt{T_\alpha/m_\alpha}$).
In order to mitigate initial transients and account for our smaller simulation domain size versus the actual experiment size, the collision frequencies are artificially increased by setting the mean free path to $50 \lambda_D$, which has been found to achieve a thermalized pre-sheath.\cite{skolar2023,bradshaw2024, bradshaw2025}
To ensure that there is a collisional plasma in the center of the domain and a collisionless plasma near the walls,
the collision frequencies from Eqs.~\ref{eq:nuee} to \ref{eq:nuie} are multiplied by a function\cite{skolar2023} $h(x)$ defined as
\begin{equation}  
	h(x) = \frac{1}{2} \bigg[ h_0(x) + h_0(-x) \bigg],
\end{equation}
where
\begin{equation}
	h_0(x) = 1 + \exp\bigg( \frac{x}{12 \lambda_D} - \frac{16}{3}   \bigg).
\end{equation}

The term $S_\alpha$ is a source term used to maintain particle conservation within the domain.\cite{li2022bohm,li2022transport,skolar2023,bradshaw2024} 
To prevent the source from impacting the sheath physics, the source term should be zero near the walls and largest in the center of the domain.
Thus, it is defined as
\begin{equation}
	S_\alpha(x,v_x) = 
	\begin{cases}
		\bigg(\displaystyle\frac{1}{L_{src}} - \displaystyle\frac{|x|}{L_{src}^2}\bigg) \Gamma_{i,out} f_{0,\alpha}(v_x) & |x| \leq L_{src} \\
		0 & \text{otherwise}
	\end{cases},
	\label{eq:source}
\end{equation} 
where $L_{src}$ is the length from the center of the domain over which the source is active,
$\Gamma_{i,out}$ is the ion particle flux that leaves the domain through both walls,
and $f_{0,\alpha}(v_x)$ is a one dimensional Maxwellian distribution whose density is unity and temperature is the initial temperature of species $\alpha$. 
Both the electron and ion particle conservation are based solely on the ion particle flux leaving the domain.
For all simulations in this work, the source length is set to $L_{src}=100\lambda_D$.

Magnetic field effects are not considered in this work. 
This can be viewed as studying the plasma-material interactions at the center of the Z-pinch, where the magnetic field would be zero.
Thus, only the electric field exists in the acceleration term in Eq.~\ref{eq:boltzmann}.
The electric field is defined as $E_x = - \partial\phi/\partial x$, 
where $\phi$ is the electric potential.
The electric potential is calculated using the Poisson equation,
\begin{equation}
	\nabla^2 \phi = - \frac{e(n_i-n_e)}{\epsilon_0},
	\label{eq:poisson}
\end{equation}
where $n_\alpha$ is the number density for species $\alpha$ and is defined as the zeroth moment of the distribution function: $n_\alpha = \int_{-\infty}^\infty f_\alpha dv$. 

Eqs.~\ref{eq:boltzmann} and \ref{eq:poisson} are solved using the code \verb|Gkeyll|.\cite{gkylDocs,juno2018discontinuous,hakim2020alias}
The equations are discretized using the discontinuous Galerkin method with an orthonormal modal serendipity basis using second order polynomials.
Eq.~\ref{eq:boltzmann} is integrated in time using a three stage third-order strong stability preserving Runge-Kutta method.\cite{gottlieb2005high}
Direct matrix inversion is used to solve Eq.~\ref{eq:poisson}.

For both ions and electrons, the velocity space uses zero flux boundary conditions.
For ions, the Eq.~\ref{eq:boltzmann} configuration space boundary condition is the perfectly absorbing wall;
any ion that enters the wall leaves the domain.
For the electrons, we run simulations with either the perfectly absorbing wall boundary condition or the IIEE model through Eq.~\ref{eq:feTotal}.
Ref.~\onlinecite{bradshaw2024} discusses the numerical implementation of Eq.~\ref{eq:feTotal} in more detail and for a more generalized case of emissions.

FuZE has used copper for the anode and graphite for the cathode.
At the moment, our model does not apply for graphite;
therefore, we use copper as the wall material for both electrodes.

The boundary conditions for Eq.~\ref{eq:poisson} are Dirichlet with the left wall set to the applied bias potential, $\phi_b$, and the right wall set to 0.
Thus, we define the anode to be at the left wall and the cathode to be at the right wall.

The ion and electron distribution functions are initialized as Maxwellians,
\begin{equation}
	f_{\alpha,0} = \frac{n_0}{ \sqrt{2\pi v_{th,\alpha,0}^2}}
	\exp \bigg( - \frac{v_x^2}{2 v_{th_{\alpha,0}}^2}   \bigg),
	\label{eq:maxwellian_IC}
\end{equation}
where $n_0$ is the initial density, 
$v_{th_{\alpha,0}}=\sqrt{T_{\alpha,0}/m_\alpha}$ is the thermal velocity,
and $T_{\alpha,0}$ is the initial temperature for species $\alpha$.
Based on experimental results from FuZE,\cite{zhang2019sustained}
the initial densities and temperatures 
are $n_{i,0}=n_{e,0} = n_0=1.1\times10^{23}\ \si{\meter^{-3}}$
and $T_{i,0}=T_{e,0} = T_0 = \SI{2}{\kilo\electronvolt}$, respectively.

The domain in configuration space is $L_D=\pm 256 \lambda_D$ with a resolution of 1024 cells.  
The velocity space for the ions and electrons span $\pm 6 v_{th_{i,0}}$ with 64 cells
and $\pm 4 v_{th_{e,0}}$ with 512 cells, respectively.
The high resolution electron velocity grid is chosen to handle the large disparities of electron energies expected in the domain. 
The Z-pinch plasma electrons have a thermal energy of \SI{2}{\kilo \electronvolt} while
the emitted electrons have energies on the order of ones to tens of \si{\electronvolt},
as shown in Fig.~\ref{f:spectrum-data-gauss-compare}. 
Thus, the electron velocity space extent and resolution are chosen to allow for reasonable computational cost of the simulations while still retaining the important physics. 

We run two sets of simulations: one without emissions to act as a control and the other with emissions to understand how IIEE affects sheath formation.
We run seven simulations in each set with varying bias potentials of 0, 1, 2, 4, 6, 8, and \SI{9}{\kilo\volt}.

Based on the temperature parameters, the bias potentials, and previous emissionless results\cite{skolar2023} from Fig.\ref{f:intro-ion-dist}, we anticipate that the anode and cathode will be bombarded with ions of about \SI{0.8}{\kilo\electronvolt} and \qtyrange{1}{10}{\kilo\electronvolt} energies, respectively.
Based on Fig.~\ref{f:model-comparison-gamma}(e), this predicts SEY of about 0.3 and 0.3 to 1 for the anode and cathode, respectively.

\section{Results} \label{s:results}

All of the results presented here are taken at a quasi-steady state with $t=\num[group-separator={,}]{10000}\omega_{pe}$ where $\omega_{pe}=\sqrt{n_0e^2/m_e\epsilon_0}$ is the plasma frequency.

\begin{figure*}[!htb] 
	\centering
	\includegraphics[width=.9\linewidth]{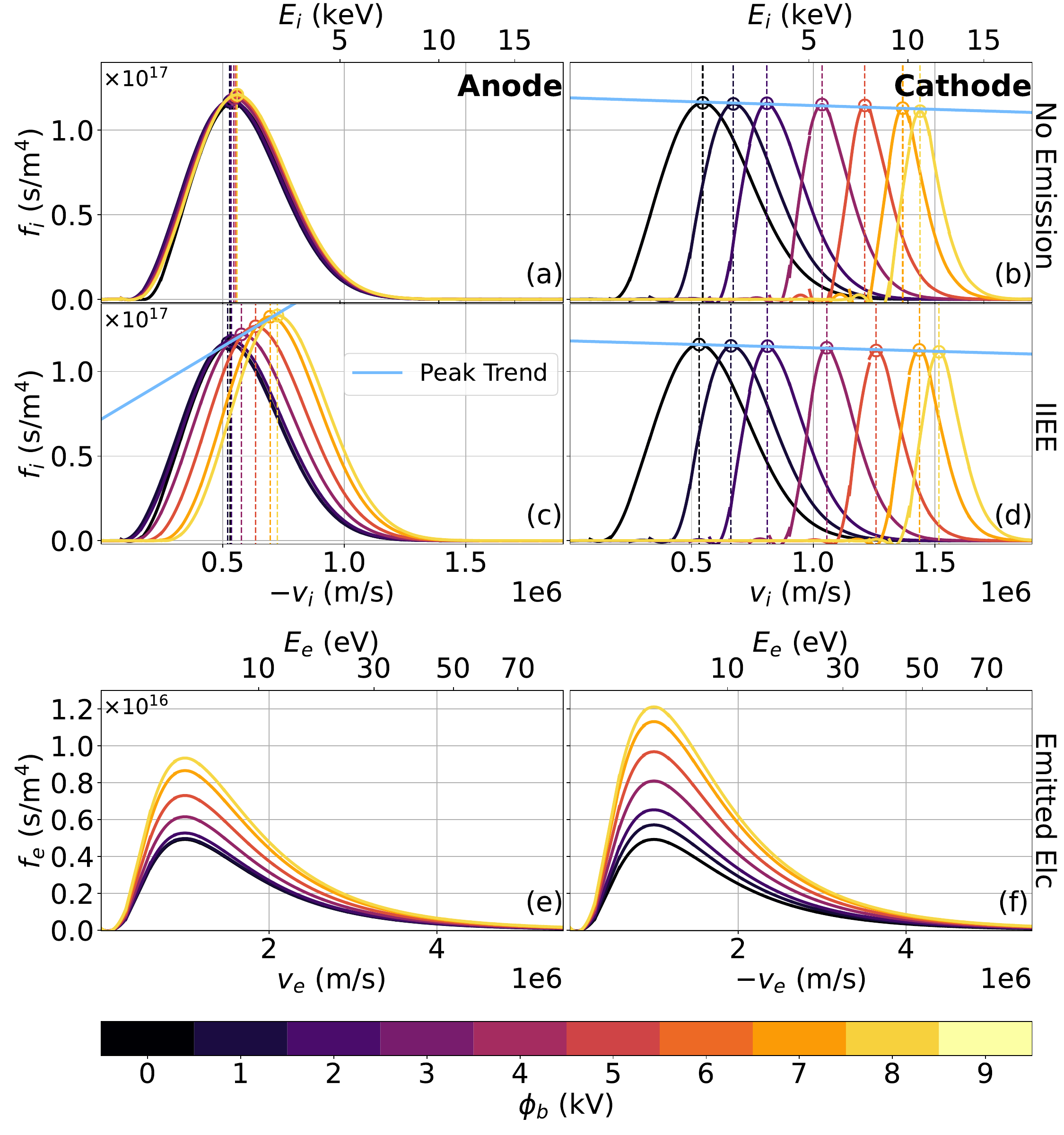}
	\caption{Plots of the impacting ion distribution function without (a-b) and with (c-d) IIEE
		at the anode (left) and cathode (right) at multiple bias potentials.
		The markers highlight the peaks of the ion distribution functions.
		The vertical dashed lines indicate the location in velocity space of the maximum value of the ion distribution function.
		Panels (e-f) show the emitted electron distribution function. 
		The trend lines (blue solid lines) show the behavior of the peak of the distribution function as the bias potential changes.
		Since the anode ion (a,c) and cathode emitted electron (f) distributions are left going based on the geometry, their velocity coordinate signs are flipped to more easily compare both electrodes.
	}
	\label{f:ion_dist_emission}
\end{figure*}

Fig.~\ref{f:ion_dist_emission} shows the impacting ion (a-d) and the emitted electron (e-f) distribution functions at each electrode.
The impacting ion distribution without emissions at the anode, Panel (a), does not show much change with bias potential, which is generally in line with Fig.~\ref{f:ion_dist_emission}(a).
The velocity of the peak initially decreases with bias potential and then increases, as denoted by the motion of the markers.
The magnitude of the distribution function also increases slightly with bias potential.

At the cathode, Panel (b), the impacting ion distribution shifts to higher energies with bias potential due to the larger potential difference.
This matches Fig.~\ref{f:intro-ion-dist}(b) but with less of a decrease in the peak value of the distribution function.
This difference is attributed to a larger configuration space domain size in these simulations allowing for better retainment of particles and energy within the domain.  
The motion of the distribution peak with bias potential follows a linear trend with velocity and is described by the line 
$f^{peak}_{i,cat,NoEm}=-\num[exponent-product=\times,output-exponent-marker=,per-mode=symbol]{4.540e9}v_i+\SI[exponent-product=\times,output-exponent-marker=,per-mode=symbol]{1.191e17}{m^{-4}s^{-1}}$.

With the inclusion of IIEE, the impacting ion distribution function at the anode, Panel (c), behaves significantly differently.
Noting the motion of the vertical dashed lines, the ion distribution function initially shifts to slightly lower energies from bias potentials of \qtyrange{0}{1}{\kilo\volt}, matching that of the no emission case.
At \SI{2}{\kilo\volt} and above, the ion distribution function becomes significantly larger valued and shifts to higher energies.
The linear trend of the peak motion is described by the line 
$f^{peak}_{i,an,IIEE}=\num[exponent-product=\times,output-exponent-marker=,per-mode=symbol]{8.600e10}v_i+\SI[exponent-product=\times,output-exponent-marker=,per-mode=symbol]{7.173e16}{m^{-4}s^{-1}}$.

At the cathode, the general trend for the IIEE case remains the same as for the emissionless case, as shown in Panel (d).
The ion distribution function shifts to higher energies and has lower peak values with increasing bias potential.
Compared to the emissionless case, however, the ion distribution function has a slightly smaller decrease compared to the emissionless case.
This is seen by examining the trend line, $f^{peak}_{i,cat,IIEE}=-\num[exponent-product=\times,output-exponent-marker=,per-mode=symbol]{4.088e9}v_i+\SI[exponent-product=\times,output-exponent-marker=,per-mode=symbol]{1.181e17}{m^{-4}s^{-1}}$,
whose slope is shallower than that without emission.
At \SI{0}{\kilo\volt} and \SI{1}{\kilo\volt}, the ion distribution function is at slightly lower energies in the IIEE case compared to the emissionless case, as noted by the leftward shift in locations of the vertical dashed lines for the IIEE cases.
Above \SI{1}{\kilo\volt}, the ion distribution function is at higher energies compared to the emissionless case.
Furthermore, as the bias potential increases, this difference between the shifts in energies for the emissionless and IIEE cases also increases.

The emitted electron distribution function at the anode, Panel (e), shows no visible change from \qtyrange{0}{1}{\kilo\volt}.
Above this, however, the emitted electron distribution function increases with bias potential.
At the cathode, Panel (f), the emitted electron distribution function continually increases with bias potential.
The increase is expected based on the ion impact energies and expected SEY from Fig.~\ref{f:model-comparison-gamma}(e).
It is not until ion impact energies of over \SI{100}{\kilo\electronvolt}, based on Fig.~\ref{f:model-comparison-gamma}(e) that the SEY is expected to start decreasing.

\begin{figure*}[!htb]
	\centering
	\includegraphics[width=\linewidth]{no-emission-cat-dists.png}
	\caption{Phase space plots of the ion (a-c) and electron (d-f) distribution functions near the cathode without emissions for varying bias potentials.
		Increasing the bias potential accelerates ions to higher energies.
		Note that the color bars for Panels (d-f) are saturated for proper comparison with the IIEE results in Figs.~\ref{f:cathode_dists_IIEE}(d-f).
	}
	\label{f:cathode_dists_NoEm}
\end{figure*}

\begin{figure*}[!htb]
	\centering
	\includegraphics[width=\linewidth]{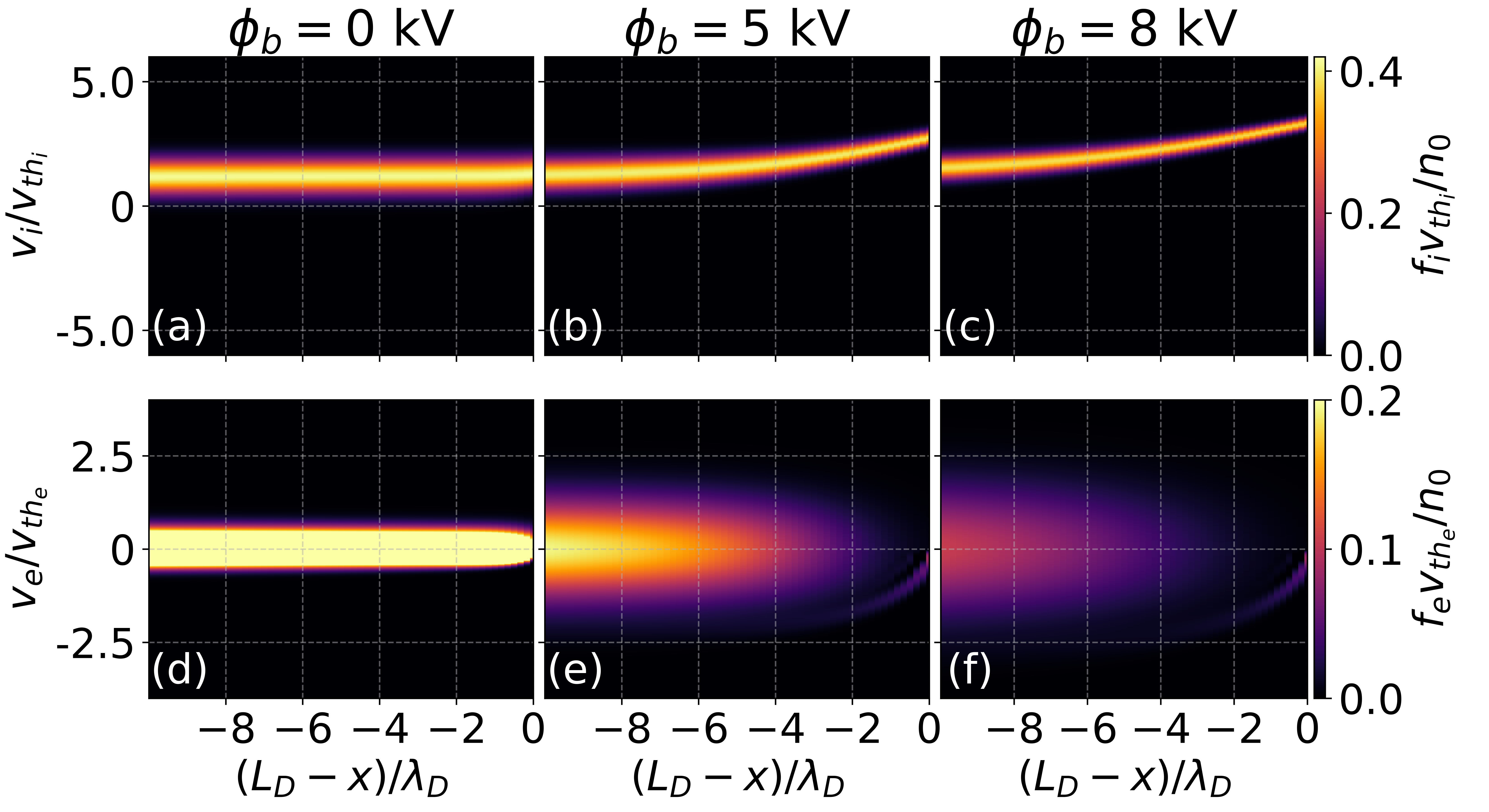}
	\caption{Same as Fig.~\ref{f:cathode_dists_NoEm} but with IIEE. 
		The emitted beam of electrons is seen accelerating to higher energies into the domain by the sheath potential.
		Note that the color bar for Panel (d) has been saturated to more easily show the electron beam in Panels (e-f).
	}
	\label{f:cathode_dists_IIEE}
\end{figure*}

Fig.~\ref{f:ion_dist_emission} only shows the behavior of the distribution functions at the electrodes.
Figs.~\ref{f:cathode_dists_NoEm} and \ref{f:cathode_dists_IIEE} show the normalized ion and electron distribution functions without and with IIEE, respectively, near the cathode for three different bias potentials.
Due to the higher bias potentials, the ions, Figs.~\ref{f:cathode_dists_NoEm}(a-c) and \ref{f:cathode_dists_IIEE}(a-c), become accelerated to higher energies, as noted by the upward shift to higher velocities closer to the cathode.
Furthermore, the ion temperature (general width of the distribution in velocity space), decreases near the cathode with increasing bias potential.
Both of these effects are stronger in the IIEE case.

Figs.~\ref{f:cathode_dists_NoEm}(d-f) show the electron distribution function near the cathode without emissions.
These are standard classical sheath results\cite{cagas2017continuum,skolar2023} showing stronger shielding of the electrons with higher bias potential.
In addition, the electron temperature (distribution width) increases near the cathode, which matches previous results.\cite{skolar2023}
The color bars are saturated for easier comparision with the IIEE cases in Figs.~\ref{f:cathode_dists_IIEE}(d-f).
The actual maximum values of the normalized electron distribution function near the cathode for the 0, 4, and \SI{9}{\kilo\volt} cases are 0.795, 0.390, and 0.355, respectively.

The story changes with emissions, as seen in Figs.~\ref{f:cathode_dists_IIEE}(d-f).
The bulk domain plasma follows a similar but weaker trend as without emissions; there is stronger shielding of the electrons with higher bias potential.
However, in these cases, an emitted electron beam is observed coming from the cathode.
Because of the disparity in the energy of Z-pinch plasma electrons in the domain (\si{\kilo\electronvolt} range) versus the energies of the emitted electrons (\qtyrange{1}{70}{\electronvolt} range), the emitted population manifests itself visually as a beam of electrons near $(L_d-x)/\lambda_D=0$.
Note that there are not sufficient emitted electrons for the \SI{0}{\kilo\volt} case, Fig.~\ref{f:cathode_dists_IIEE}(d), to see such a beam.

The emitted electrons are accelerated to higher energies into the domain because they are traveling against the sheath electric field.
Higher bias potentials accelerate the electrons to even higher energies, as shown by the downward shift in location of the emitted electron beam between Figs.~\ref{f:cathode_dists_IIEE}(e) and \ref{f:cathode_dists_IIEE}(f).
These high energy electrons collide with the electrons in the domain causing an increase in the electron temperature.
This is most readily noted by examining the width of the \SI{0}{\kilo\volt} case compared to the \SI{5}{\kilo\volt} and \SI{9}{\kilo\volt} cases.
This effect is also seen in the no emission cases, but to a lesser degree.

The sheath potential barrier significantly reduces the electron population near the cathode with the effect becoming stronger with increasing bias potential.
To counter this effect while also showing the differences with the \SI{0}{\kilo\volt} case regarding the emitted electron beam and the temperature, we choose to saturate the color bar for Fig.~\ref{f:cathode_dists_IIEE}(d) with a maximum color value at 0.26.
The actual maximum values of the normalized electron distribution function near the cathode for \SI{0}{\kilo\volt} IIEE case is 0.804.

\begin{figure}[!htb]
	\centering
	\includegraphics[width=\linewidth]{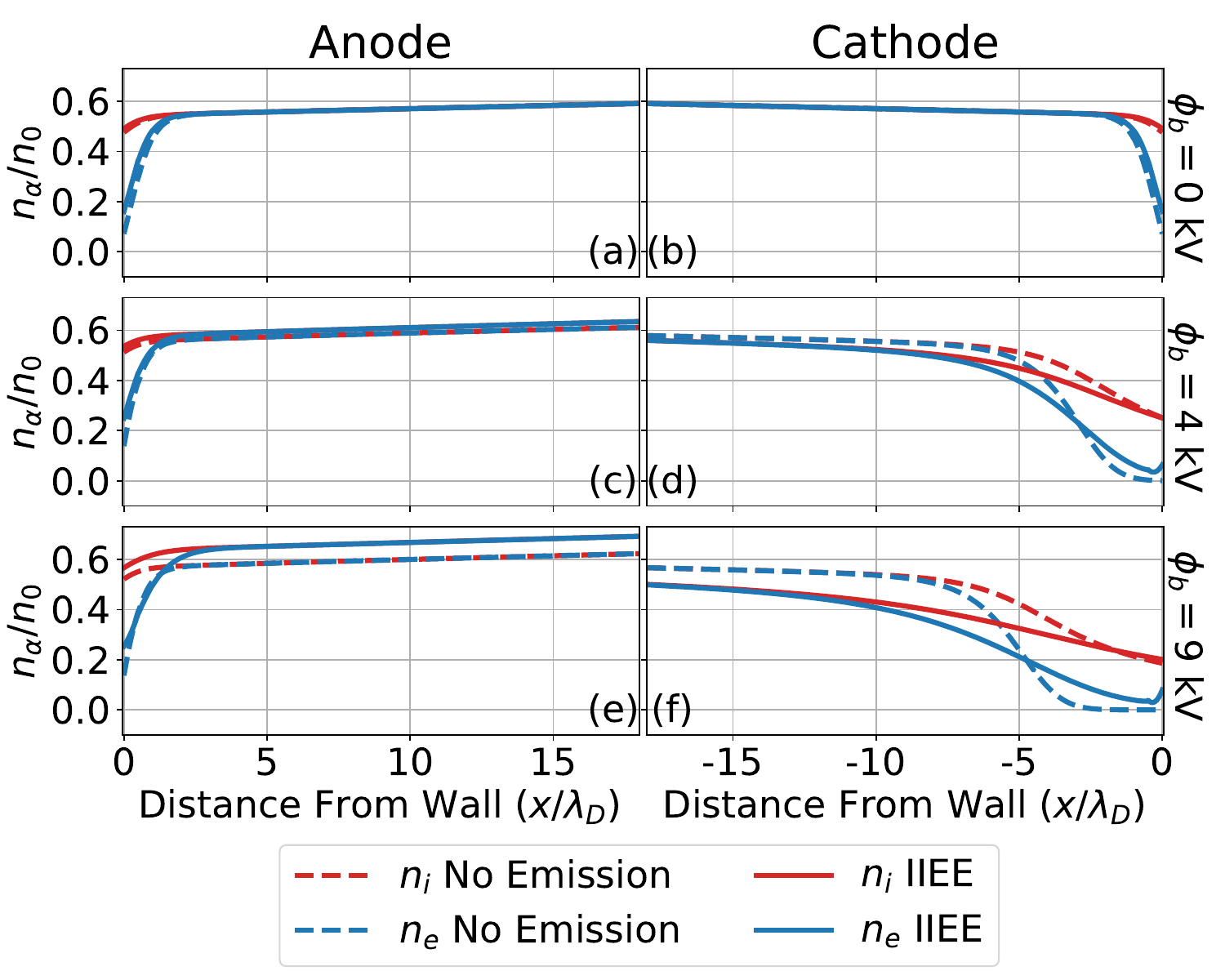}
	\caption{Plots of ion (red) and electron (blue) densities with (solid lines) and without (dashed lines) emissions near the anode (a,c,e) and the cathode (b,d,f) for the \SI{0}{\kilo\volt} (a,b), \SI{4}{\kilo\volt} (c,d), and \SI{9}{\kilo\volt} (e,f) bias potential cases.
	Near the cathode, the electron density becomes non-monotonic at higher bias potentials due to the emissions.}
	\label{f:compare_dens}
\end{figure}

Fig.~\ref{f:compare_dens} shows the ion (red) and electron (blue) densities with (solid lines) and without (dashed lines) emissions near the anode and the cathode for the 0, 4, and \SI{9}{\kilo\volt} cases.
All of the emissionless cases follow the generally expected monotonic density decrease near the wall with lower densities at the cathode for higher bias potentials.\cite{stangeby2000plasma,skolar2023}
As should be expected, the ion and electron densities for every case (with and without emissions) converge to maintain quasineutrality toward the center of the domain.

We focus first on the anode.
At \SI{0}{\kilo\volt} and \SI{4}{\kilo\volt}, the ion and electron densities are slightly larger with emissions compared to without IIEE.
At \SI{9}{\kilo\volt}, the ion density is noticeably larger with IIEE than without. 
The electron density is greater with IIEE immediately near the anode but is smaller further away before converging.

Now focusing on the cathode, the \SI{0}{\kilo\volt} result is the same as that of the anode due to symmetry.
At \SI{4}{\kilo\volt}, the ion density for the IIEE case is the same at the wall but lower further away.
At \SI{9}{\kilo\volt}, the ion density is greater for the IIEE case immediately near the wall but is less further away.
The electron density at \SI{4}{\kilo\volt} is greater near the wall for the IIEE case and is less further away.
This effect becomes stronger for the \SI{9}{\kilo\volt} case.

The increase in electron density near the electrodes with emissions compared to without emissions is the direct result of emitted electrons entering the domain from the wall.
In addition, the higher bias potential cases show the electron density has a non-monotonic increase near the cathode, which is not typical with standard emissionless sheath simulations.

\begin{figure}[!htb] 
	\centering
	\includegraphics[width=\linewidth]{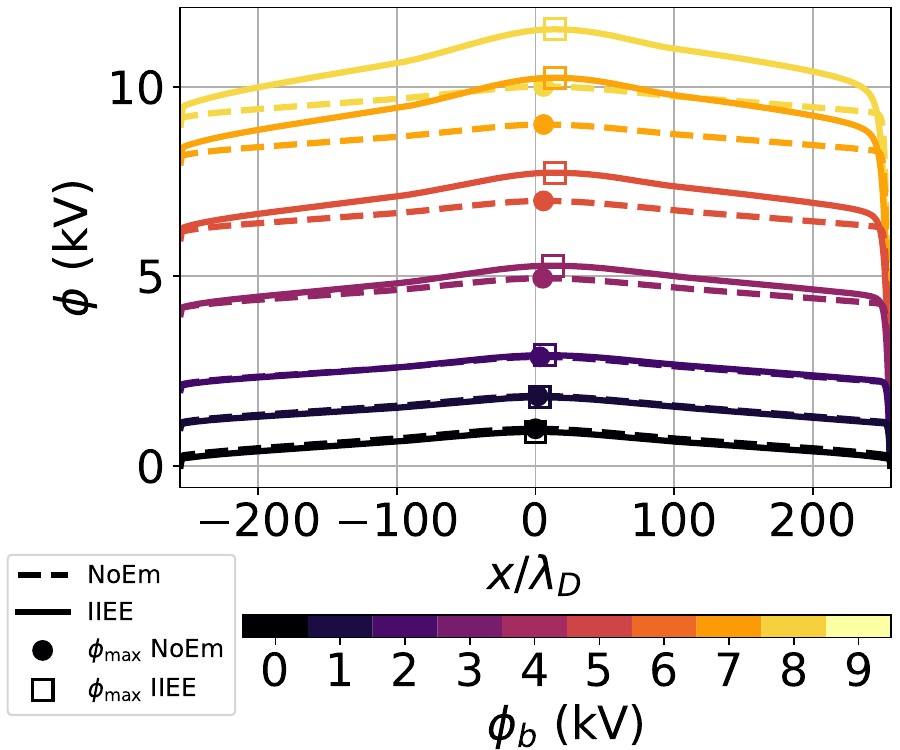}
	\caption{Plot of the electric potential profile for various bias potentials.
		The dashed and solid lines correspond to the no emission and IIEE cases, respectively.
		The circles and squares correspond to the maximum potentials for the no emission and IIEE cases, respectively.
		For all cases, the electric potential follows a monotonic profile with a plasma potential greater than the bias potential.
	}
	\label{f:potential}
\end{figure}

Despite the non-monotonicity, the electron density is always lower than the ion density for all bias potentials. 
This suggests the formation of a classical sheath with a monotonic potential profile, as shown in Fig.~\ref{f:potential}.
As bias potential increases, the entire profile increases for both the no emission (dashed lines) and IIEE (solid lines) cases, as expected from theory.\cite{stangeby2000plasma}
For the \SI{0}{\kilo\volt} and \SI{1}{\kilo\volt} cases, the electric potential is slightly lower for the IIEE cases compared to the no emission cases.
At a bias potential of \SI{2}{\kilo\volt}, the electric potential is slightly greater near the cathode (right) and slightly lower near the anode (left) for the IIEE case versus the no emission case.
For bias potentials above \SI{2}{\kilo\volt}, the electric potential for the IIEE case becomes increasingly greater than the emissionless case throughout the entire domain.

The circles and squares in Fig.~\ref{f:potential} correspond to the maximum potential, which we define as the plasma potential, for the no emission and IIEE cases, respectively.
At \SI{0}{\kilo\volt}, the maximum potential is in the center of the domain for both cases due to symmetry.
As the bias potential increases, the spatial location of the maximum potential generally shifts closer to the cathode (to the right) for both cases due to the natural asymmetry of the potential bias between the electrodes.
This shift is stronger for the IIEE cases versus the emissionless cases.

\begin{figure}[!htb]
	\centering
	\includegraphics[width=\linewidth]{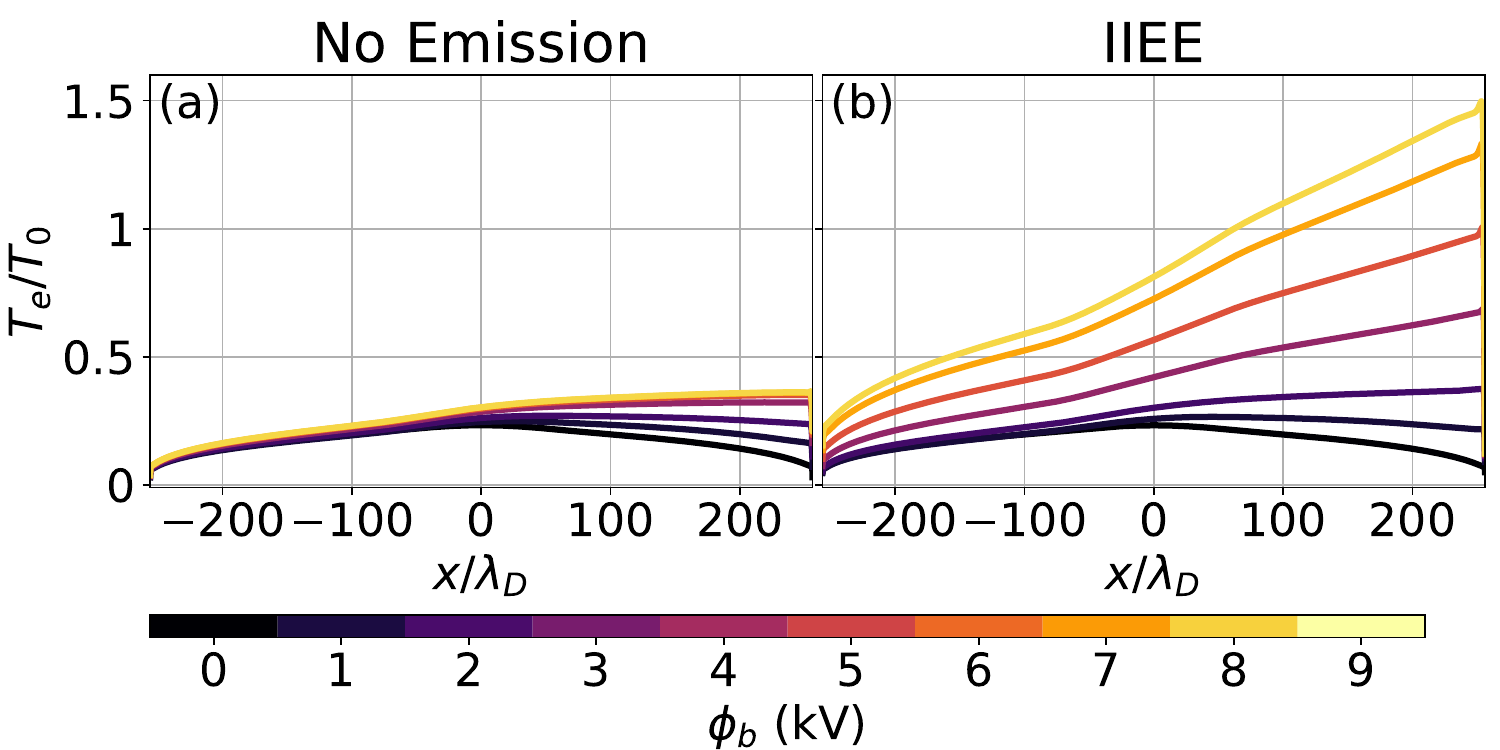}
	\caption{Plots of the electron temperature profiles without (a) and with (b) emissions.
		The emitted electrons accelerate to higher energies heating the domain electrons.}
	\label{f:temp_profile}
\end{figure}

With increasing plasma potential, the emitted electrons are accelerated to higher energies.
These high energy emitted electrons collide with the electrons in the domain transferring some of their energy.
This results in higher temperatures throughout the domain, as shown in Fig.~\ref{f:temp_profile}.
The temperature is calculated using
\begin{equation}
	T_\alpha = m_\alpha \bigg( \frac{M_{2,\alpha}}{n_\alpha} - u_\alpha^2\bigg),
	\label{eq:temp}
\end{equation}
where $M_{2,\alpha}=\int_{-\infty}^\infty v^2 f_\alpha dv$ is the second moment of the distribution function
and $u_\alpha$ is the drift velocity found using the density and the particle flux.
The particle flux is the first moment of the distribution function: $\Gamma_\alpha=n_\alpha u_\alpha = \int_{-\infty}^\infty v f_\alpha dv$.

Without emissions and as bias potential increases, the electron temperature decreases near the anode and increases near the cathode, as shown in Fig.~\ref{f:temp_profile}(a).
With emissions, however, the emitted electrons add energy to the domain and increase the temperature significantly while retaining the general asymmetry between the electrodes, as shown in Panel (b).
This is because the emitted electrons from the cathode lose energy due to collisions as they travel further into the domain (closer to the anode). 
Since the emitted electrons are accelerated to higher energies with higher bias potential, the temperature increase also increases with bias potential. 

\begin{figure}[!htb] 
	\centering
	\includegraphics[width=\linewidth]{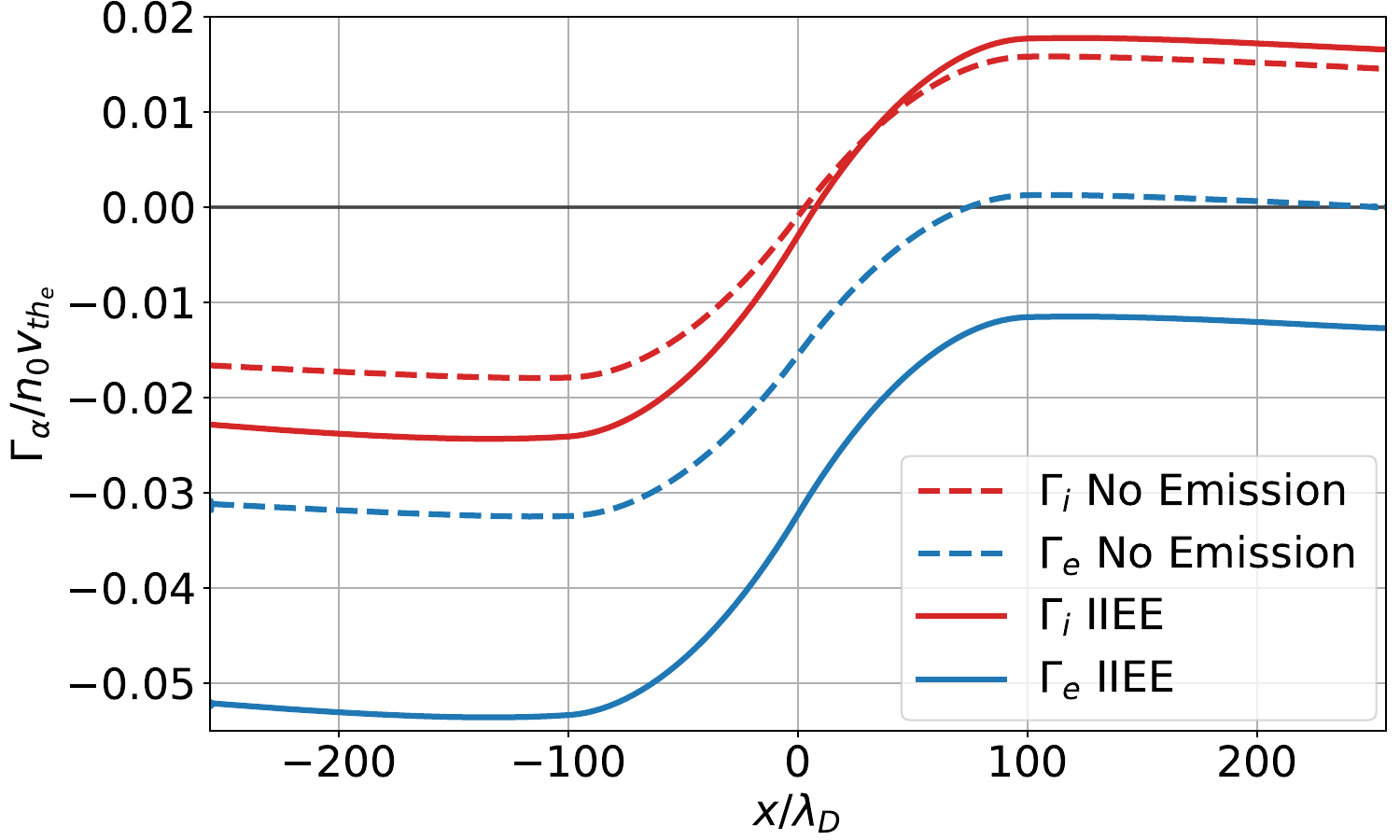}
	\caption{Plot of the ion (red) and electron (blue) particle fluxes with (solid lines) and without (dashed lines) emissions. 
		The inclusion of emissions allows the electron particle flux at the cathode to be negative, thereby increasing the current in the domain.
		Note that all particle fluxes are normalized using the electron thermal velocity for proper comparison.} 
	\label{f:niui_profile_8kV}
\end{figure}

The emitted electrons allow there to be a particle flux where there otherwise would be none.
Fig.~\ref{f:niui_profile_8kV} shows the normalized particle flux profiles over the entire domain for the maximum emission case (\SI{9}{\kilo\volt}).
With IIEE, the ion particle flux is slightly larger at the cathode but significantly larger at the anode.
In addition, the emissions induce an asymmetry in the ion particle flux with the zero flux point shifted closer to the more emissive cathode.

Without emissions, the electron particle flux tends to zero near the cathode for high bias potentials.
This saturation is a consequence of the absorbing wall boundary condition.
Fluxes can only go into the wall or be zero;
they cannot come out of the wall.\cite{skolar2023}
With emissions, however, electrons are emitted from the wall into the domain;	
in other words, the electron particle flux comes out of the wall and is negative at the cathode.
Thus, since more electrons are emitted from the cathode compared to the anode, the entire electron particle flux profile is shifted in the negative direction.

The ion and electron particle fluxes follow the same profile shape such that the current density, defined as $j=e(n_iu_i-n_eu_e)$, in the entire domain is constant with and without IIEE.

Without emissions, the ions and electrons provide equal contributions to the current at $x/\lambda_D=29.5$.
Below this, the current is driven primarily by the electrons;
above this, the current is driven primarily by the ions with the electron contribution going to zero at the cathode.

With emissions, the current density increases because the electron particle flux becomes more negative.
Therefore, the electrons contribute more to the current over a larger stretch of the domain with the equal contribution point now shifted to $x/\lambda_D=63.3$.
Furthermore, the ions change from entirely driving the cathode current in the emissionless case to only contributing to 56.5\% of the cathode current with emissions.

\begin{figure}[!htb]
	\centering
]	\includegraphics[width=\linewidth]{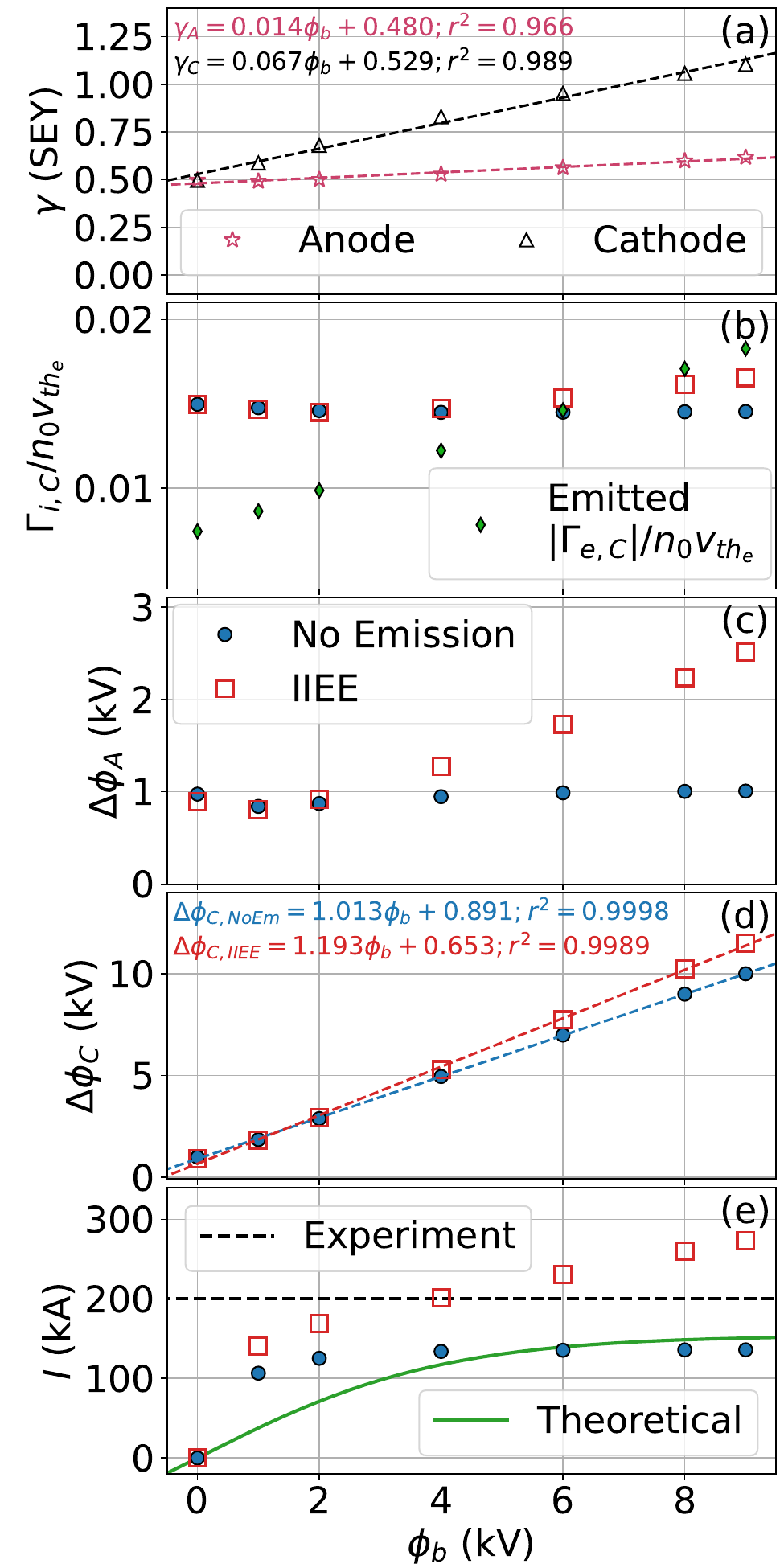}
	\caption{Plots of the SEY (a),
		normalized ion particle flux at the cathode (b),
		anode potential difference (c),
		cathode potential difference (d),
		and current (e) 
		as a function of the bias potential.
		Panel (b) also shows the normalized emitted electron particle flux at the cathode.
		Increasing the bias potential increases the emissions at both walls resulting in a larger current that, for some bias potentials, better matches experiment.\cite{zhang2019sustained} 
	}
	\label{f:gamma_phi_I}
\end{figure}

To fully understand the emissions, we calculate the SEY ($\gamma$).
Since only the impacting ion particle flux and the emitted electron particle flux are used for Eq.~\ref{eq:gamma}, we must take the one sided first moment of the corresponding distribution functions from Figs.~\ref{f:ion_dist_emission}(c-f). 
In other words, we must find the left or right going particle fluxes of the appropriate distribution functions: $\Gamma_{\alpha,L} = \int_{-\infty}^0 v f_\alpha dv$ or $\Gamma_{\alpha,R} = \int_{0}^\infty v f_\alpha dv$.
We calculate the SEY at the anode by using the right going emitted electron particle flux and the left going impacting ion particle flux for Eq.~\ref{eq:gamma}.
Similarly, the SEY at the cathode is found by using the left going emitted electron particle flux and the right going impacting ion particle flux.

Fig~\ref{f:gamma_phi_I}(a) shows that the SEY at the cathode (black triangles) increases significantly with bias potential.
This is the expected result based on the shift of the ions to higher energies at the cathode, Fig.~\ref{f:ion_dist_emission}(d), in conjunction with how $\gamma$ increases with ion energy in this energy regime, as shown in Fig.~\ref{f:model-comparison-gamma}(e).
Except for the \SI{1}{\kilo\volt} case, the anode (pink stars) shows a slight increase in the SEY with bias potential.
While it is not easily visible, the SEY decreases to 0.491 for the \SI{1}{\kilo\volt} case from 0.497 for the \SI{0}{\kilo\volt} case.
This is in agreement with the impacting ion distribution function at the anode shifting to slightly lower energies from \qtyrange{0}{1}{\kilo\volt} in Fig.~\ref{f:ion_dist_emission}(c);
the peak value of the ion distribution function at the anode for the \SI{0}{\kilo\volt} and \SI{1}{\kilo\volt} IIEE cases are located at velocities of \SI[exponent-product=\times,output-exponent-marker=,per-mode=symbol]{5.306e5}{\meter\per\second} and
\SI[exponent-product=\times,output-exponent-marker=,per-mode=symbol]{5.211e5}{\meter\per\second}, respectively.
This pattern exactly matches that of the SEY at the anode.

The SEY at both the anode and the cathode follow a linear relationship.
The slope is dependent on how the ion energies at the electrodes change with bias potential.
Therefore, the slope is larger at the cathode than at the anode.

Focusing on the SEY at the cathode, it is clear from Fig.~\ref{f:ion_dist_emission}(f) that the emitted electron particle flux increases by the increasing electron distribution function and is further shown in Fig.~\ref{f:gamma_phi_I}(b) (green diamonds).
However, the SEY is also dependent on the impacting ion particle flux.
Without emissions, the ion particle flux (blue circles) decreases slightly to an asymptote with increasing bias potential.
With emissions (red squares), the ion particle flux follows the same patterns at low bias potentials, but significantly increases at high bias potentials. 
At \SI{8}{\kilo\volt}, the emitted electron particle flux is larger than the ion particle flux corresponding to an SEY greater than 1, as also shown in Fig.~\ref{f:gamma_phi_I}(a).

The SEY behavior can be explained by the potential differences between the plasma and the electrodes. The plasma potential, $\phi_p$, is taken as the maximum electric potential within the domain, as shown by the markers in Fig.~\ref{f:potential}.

Fig.~\ref{f:gamma_phi_I}(c) shows the anode potential difference, $\Delta \phi_A = \phi_p - \phi_b$. 
Without emissions, the anode potential difference decreases to an asymptotic limit with increasing bias potential.
With emissions, the potential difference is lower for the 0 and \SI{1}{\kilo\volt} bias potential cases compared to the emissionless case;
this is the same pattern as what is found in the anode SEY annd the shifts in velocities of the distribution function at the cathode.
The lower SEY in this region is due to the smaller potential difference causing decreased ion energies at the anode, as shown in Fig.~\ref{f:ion_dist_emission}(c).

The cathode potential difference ($\Delta \phi_C = \phi_p$), Fig.~\ref{f:gamma_phi_I}(d), generally linearly increases with bias potential for both the emission and emissionless cases.
Beginning at \SI{2}{\kilo\volt}, the potential difference with emissions is greater than that without emissions resulting in a larger slope for the trend line.
This leads to the ions accelerating to higher energies resulting in higher SEY.

The change in the SEY with bias potential leads to changes in the ion and electron particle fluxes which changes the current density.
Assuming the current density is uniformly distributed across the cylindrical Z-pinch, the current is approximated by $I = j \pi a^2$ where $a$ is the pinch radius, which for FuZE is \SI{3}{\milli\meter}.\cite{zhang2019sustained}
Fig.~\ref{f:gamma_phi_I}(e) compares the current in the domain to emissionless theory\cite{stangeby2000plasma} (solid green line) and experiment\cite{zhang2019sustained} (dashed black line).
The current in the domain is taken at $x=0$;
note that because the current density is generally spatially constant within the domain.
Assuming no emissions, the theoretical current is calculated using\cite{stangeby2000plasma}
\begin{equation}
	I_{theo} = \frac{1}{2} e n_0 c_s \pi a^2 \Bigg[ 1 - \exp\bigg(\frac{e [\phi_{p_0}-\phi_p]}{T_{e,0}}\bigg)  \bigg],
\end{equation}
where $c_s= \sqrt{(T_{i,0}+T_{e,0})/m_i}$ is the isothermal ion acoustic speed, $\phi_{p_0}$ is the theoretical plasma potential with zero bias potential defined as\cite{stangeby2000plasma}
\begin{equation} 
	\phi_{p_0} = - \frac{1}{2}  \frac{T_{e,0}}{e}
	\ln \Bigg[ \bigg( 2 \pi \frac{m_e}{m_i} \bigg) \bigg( 1 + \frac{T_{i,0}}{T_{e,0}} \bigg) \Bigg],
	\label{eq:phip0_theo}
\end{equation}
and $\phi_p$ is the theoretical plasma potential defined as\cite{stangeby2000plasma}
\begin{equation}
	\phi_{p} = \phi_b 
	- \frac{T_{e,0}}{e} \ln \left[ 
	\frac{2 \exp\Big(- \displaystyle\frac{e \phi_{p_0}}{T_{e,0}}\Big)}{1 + 
		\exp\Big(- \displaystyle\frac{e\phi_b}{T_{e,0}}\Big)} 
	\right].
	\label{eq:phip_theo}
\end{equation}

The emissionless theory predicts that the current saturates to an asymptotic limit of $e n_0 c_s \pi a^2/2$ as the bias potential tends to $\infty$.
The emissionless simulation results reach a lower asymptotic limit of about \SI{130}{\kilo\ampere} at lower bias potentials matching previous work.\cite{skolar2023}
With the inclusion of IIEE, the current in the domain increases with bias potential reaching about \SI{273}{\kilo\ampere} at \SI{9}{\kilo\volt}.
The current most closely matches experiment at \SI{4}{\kilo\volt} with a value of \SI{201}{\kilo\ampere}.
Therefore, the disparity between the theory and experiment can be partially explained by the inclusion of IIEE.
Note that the experiment uses graphite for the cathode\cite{zhang2019sustained} whereas we use copper.
In addition, the magnetic field affects are expected to lower the SEY and thus lower the current.
Therefore, while not an exact comparison, this work still shows clearly that the inclusion of IIEE increases the current in the entire domain.

\section{Discussion}
\label{s:discussion}

Previous work on emissive sheaths have theorized the possibility of space-charge limited (SCL)\cite{hobbs1967SCL,schwager1993SCL,mcadams2012} and inverse sheaths\cite{campanell2013negative,campanell2016} when the SEY is greater than 1.
In our simulations, the SEY is greater than 1 starting at a bias potential of \SI{8}{\kilo\volt}.
However, based on the positive space charge in the sheath (Fig.~\ref{f:compare_dens}) and the monotonic potential profiles (Fig.~\ref{f:potential}), the type of sheath that forms for all bias potentials is a classical sheath.

\begin{figure}[!htb]
	\centering
	\includegraphics[width=.7\linewidth]{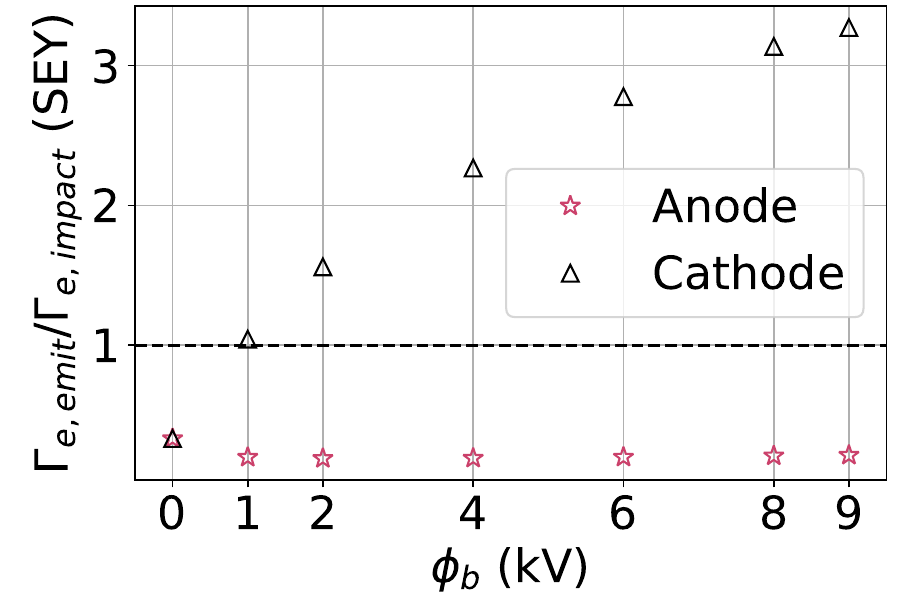}
	\caption{Plot of SEY at the anode (pink stars) and the cathode (black  triangles) using the impacting electron particle flux as opposed to the impacting ion particle flux.
		Starting at a bias potential of \SI{1}{\kilo\volt}, the electron-based SEY is greater than 1 at the cathode.
	}
	\label{f:gamma_e}
\end{figure}

Previous works are primarily focused on electron induced or thermionic electron emission. 
Therefore, they use the impacting electron particle flux for their SEY calculations as opposed to the impacting ion particle flux, as in our case.
Fig.~\ref{f:gamma_e} shows the SEY if calculated using the impacting electron particle flux from our IIEE simulations. 
At the anode, the electron-based SEY remains below 1 for all bias potentials.
At the cathode, however, the electron-based SEY is greater than 1 starting at \SI{1}{\kilo\volt}.
Despite this, a classical sheath forms for all cases.

To understand further, we will examine the theoretical space charge limits that may cause the SCL or inverse sheaths.
The theoretical emitted electron particle flux saturation limit above causes a virtual cathode or SCL sheath is\cite{mcadams2012}
\begin{equation}
	\Gamma_{e,emit,sat} = (n^{se}_i - n^{se}_e) \sqrt{ \frac{2 e (\phi^{se} + E_{e,emit})}{m_e}},
	\label{eq:SCL_limit}
\end{equation}
where the superscript $se$ denotes the values at the sheath edge and $E_{e,emit}$ is the emitted electron energy in \si{\electronvolt}.
We use the electron energy of peak emission for $E_{e,emit}$ which is $E_0$ from Eq.~\ref{eq:gauss};
for copper, this is \SI{2.807}{\electronvolt}.
For all of our simulations, the saturation limit is on the order of \SI[exponent-product=\times,output-exponent-marker=,per-mode=symbol]{e29}{\meter^{-2}\second^{-1}} or larger
whereas the largest emitted electron particle flux is on the order of \SI[exponent-product=\times,output-exponent-marker=,per-mode=symbol]{e28}{\meter^{-2}\second^{-1}}.
Therefore, we would not expect an SCL sheath to form.

The other mechanism by which a classical sheath may transform to a SCL or inverse sheath with increasing emissions is through backflow current saturation based on collisional effects.\cite{campanell2025}
The space charge limit decreases for highly collisional plasmas based on the ratio of the electron mean free path to the domain length (lower is more collisional).\cite{campanell2025}
For our case, this ratio is 0.0977 which seems to be large enough to allow for a classical sheath to form.
This mechanism has only recently been discovered and has only been studied in the context of low temperature plasmas and future work should examine its application to higher energy plasmas.

In addition, it is unlikely for an inverse sheath to form because they require a cold ion population with ion-neutral collisions.\cite{campanell2013negative,campanell2016}
Such a population would accumulate overtime transitioning an SCL sheath to an inverse sheath. 
A cold ion population is unlikely to exist in the \si{\kilo\electronvolt} energy Z-pinch plasma, especially near the cathode where the ions are accelerated to even higher energies into the wall.

There are, however, behaviors in the plasma potential that match expected behavior for SCL and inverse sheaths.
In the \SI{0}{\kilo\volt} case, the plasma potential with emissions is lower than that without emissions, as shown in Fig.~\ref{f:gamma_phi_I}(c).
This is the expected result as the emitted electrons decrease the positive space charge in the sheath region causing a lower plasma potential.\cite{bradshaw2024}
At a bias potential of \SI{1}{\kilo\volt}, the increase in emissions, Fig.~\ref{f:gamma_phi_I}(b), further decreases the positive space charge and further reduces the plasma potential for the IIEE case compared to the emissionless case.
However, counter to typical SCL and inverse sheath transition behavior, this trend changes above \SI{1}{\kilo\volt} where the plasma potential is larger with emissions than without despite the continued increase in electron emissions.

As with other emissive sheaths, the emitted electrons are accelerated to higher energies by the cathode potential difference, as shown in Figs.~\ref{f:cathode_dists_IIEE}(e-f).
This effect is exacerbated as bias potential increases.
Therefore, at higher bias potentials, the accelerated emitted electrons collide with the electrons in the domain increasing the electron temperature, as shown in Fig.~\ref{f:temp_profile}(b).

\begin{figure}[!htb]
	\centering
	\includegraphics[width=.7\linewidth]{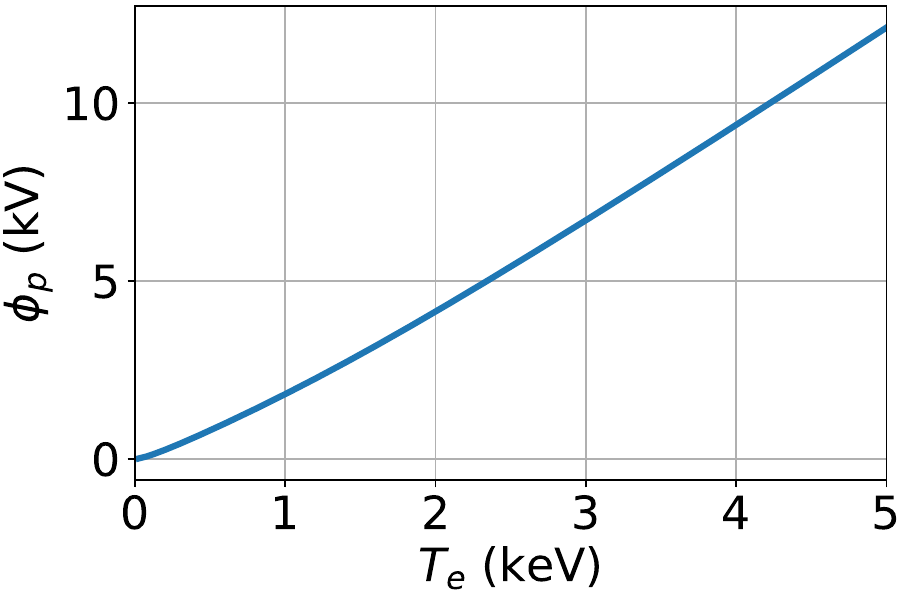}
	\caption{Plot showing how the theoretical plasma potential (Eq.~\ref{eq:phip}) increases with electron temperature.
		For this calculation, the ion temperature is \SI{1}{\kilo\electronvolt}, the bias potential is \SI{4}{\kilo\volt}, and the proton-electron mass ratio is used.
	}
	\label{f:phip}
\end{figure}

The plasma potential is highly dependent on the electron temperature.
The theoretical plasma potential in Eq.~\ref{eq:phip_theo} uses the initial temperatures and assumes an isothermal plasma.
Fig.~\ref{f:temp_profile} shows that the isothermal assumption is false and that the plasma temperature deviates signficantly from the initial temperature.
If using the actual temperature instead of the initial temperature and substituting Eq.~\ref{eq:phip0_theo} into Eq.~\ref{eq:phip_theo}, the theoretical plasma potential, 
\begin{equation}
	\phi_p = \phi_b - \frac{T_e}{e} 
	\ln \left[
	\frac{ \sqrt{\Big( 8 \pi \frac{m_e}{m_i}\Big) \Big( 1 + \frac{T_i}{T_e} \Big)} }{1 + \exp \Big( - \frac{e \phi_b}{T_e} \Big)}
	\right],
	\label{eq:phip}
\end{equation}
becomes a function of the bias potential, ion and electron mass, and ion and electron temperature.
All of our cases use a proton-electron mass ratio.
The ion temperature slightly decreases with bias potential, as shown by the smaller distribution widths in Figs.~\ref{f:cathode_dists_IIEE}(c-d).
The electron temperature significantly increases with bias potential, as shown in Fig.~\ref{f:temp_profile}(b).
The term in the logarithm will always be negative for realistic plasma values;
therefore, in conjunction with its associated negative sign, it acts as an additive term on top of the bias potential.
This term is scaled by the electron temperature;
therefore, a higher electron temperature results in a higher plasma potential, as shown by Fig.~\ref{f:phip}.

Consequently, at 0 and \SI{1}{\kilo\volt}, the electrons are not accelerated to high enough energies to sufficiently heat the plasma to increase the plasma potential.
At \SI{2}{\kilo\volt} and above, the increased electron temperature is significant enough to increase the plasma potential countering the expected drop due to more emissions.

For simulations with larger domain sizes, the effect of the electron heating is expected to be weaker.
Larger domain simulations effectively have a larger thermal mass;
there are more electrons in the domain but the same amount of electrons emitted from the wall.
Therefore, the overall heating is reduced. 
In addition, this behavior may be different for plasma models with better conservation of energy as much of the electron temperature is lost to the wall.
Note that the choice was made in our plasma model to conserve particles at the cost of energy, as shown in Eqs.~\ref{eq:boltzmann} and \ref{eq:source}.
Future work may consider implementing a power inflow boundary condition to better account for the high density and temperatures in the core plasma.\cite{li2023}

Despite the formation of only a classical sheath, it is important to consider what may happen if the emissions were sufficiently large enough to induce an inverse sheath.
The potential well of the classical sheath accelerates ions to higher energies at the wall, which increases the SEY in this energy regime per Fig.~\ref{f:model-comparison-gamma}.
If an SCL or inverse sheath were to form due to sufficient IIEE, the sheath potential will be such that the ions are slowed, reducing their impact energies at the wall. 
This would significantly reduce the electron emission and a classical sheath will likely form again.
At which point, the ions are accelerated into the wall increasing emissions and the cycle continues. 
A priori, it is unclear if the plasma would remain dynamically changing between classical, SCL, and inverse sheaths or if some sort of equilibrium would be reached.

It may be possible for a more dynamic sheath to form if we can increase the SEY at low bias potentials where the emissive electron heating plays a smaller role (less than \SI{2}{\kilo\volt}).
Without sufficient heating, a strong SEY may continue to reduce the plasma potential to form SCL or inverse sheaths.
We examine several factors that may impact the SEY.

``Dirty" or oxidized materials have higher IIEE SEY\cite{hasselkamp1992book,bogaerts2002,bradshaw2025} compared to ``clean" materials.
This effect is seen in Fig.~\ref{f:model-comparison-gamma}(c).
Some of the markers\cite{aarset1954,foti1974,beck1975,thornton1977,borovsky1988} appear to follow a curve that is higher than the other data points.
This is because their aluminum samples were likely not properly cleaned. 
This is evident because the Borovsky et al. (1988)\cite{borovsky1988} dataset (bottom-filled circles) is for aluminum oxide and is on that same upper curve.
Realistically, the electrodes will oxidize over time and increase the SEY.
Therefore, depending on the material and oxidation parameters, the SEY increase may be significant enough to push the low bias potential cases towards SCL or inverse sheath development.

Another method of increasing the SEY at lower bias potentials is to have higher ion energies at the wall.
This can be achieved by having a higher initial ion temperature.
At present, FuZE has an ion temperature of about \SI{2}{\kilo\electronvolt} but has not yet achieved breakeven.\cite{zhang2019sustained}
Breakeven is defined as $Q=1$ where $Q$ is the ratio of the output energy to the input energy.
To get to breakeven fusion energy, Z-pinch scaling studies predict ion temperatures of about \SI{5}{\kilo\electronvolt}.\cite{shumlak2020z}
But for realistic power plants, $Q$ must be much higher which means the ion temperature must be higher.
For $Q=10$, the ion energy is about \SI{11}{\kilo\electronvolt}.\cite{shumlak2020z}
Therefore, future Z-pinch regimes may have sufficient ion energies to increase the SEY at low bias potentials and induce SCL or inverse sheaths.

The effect of electron induced secondary electron emission may also increase the total SEY.\cite{furman2002probabilistic,bradshaw2024}
The electron particle flux into the anode increases with bias potential\cite{skolar2023} and even moreso with emissions, as shown in Fig.~\ref{f:niui_profile_8kV}.
The increase in electron emissions due to electron impacts at the anode may result in global changes to the current.

While the factors above may increase the SEY, the inclusion of a magnetic field may decrease the SEY.
The simulations in this work are in 1X-1V and neglect magnetic field effects. 
Based on a simplified Z-pinch configuration, the current in the axial direction (in the $x$ direction into the wall in our framework) induces a magnetic field in the azimuthal direction, which is on the plane parallel to the electrode.
This magnetic field causes particle gyrations along the set of planes perpendicular to the electrode.
Therefore, it is possible for an emitted electron to leave the electrode, undergo a partial orbit about the azimuthal magnetic field, collide with the same electrode, and leave the domain.
If electron induced electron emissions are considered, this may result in a sort of cascading effect with rich physics to explore within an electron gyroradius from the wall.
In addition, the effective SEY will decrease due to fewer electrons being able to accelerate into the domain. 
The amount of SEY reduction is determined by the interplay of the magnetic field strength, emitted electron energy, and the cathode potential difference.

The SEY will also change by improving the IIEE model.\cite{schou1980}
Our 1X-1V simulations assume that the emitted electrons travel only along the $x$ axis.
In reality, there is a spread of directions the emitted electrons will travel determined by a cosine distribution.\cite{schou1980}
The emitted electron energy spectrum in Eq.~\ref{eq:dgdE} is the result of an integral over the entire angular emission spectrum.\cite{schou1980}
The effect of considering the full angular dependence and additional spatial dimensions may affect the plasma potential and therefore modify the SEY. 
A numerical model has been developed for the angular dependence within our framework of using the Gaussian approximation for continuum kinetic simulations.\cite{bradshaw2024}

In addition, the model significantly underpredicts the SEY for tungsten and lithium, as shown in Figs.~\ref{f:model-comparison-gamma}(a-b).
Both of these materials are highly relevant for nuclear fusion applications. 
If the ion stopping powers use TRIM\cite{srim2010} instead of Eq.~\ref{eq:low-energy-ion-stopping-power}, the results improve significantly, especially for tungsten.
Therefore, better estimates of the low energy ion stopping power\cite{srim2010,haque2019} will improve the SEY.

Recently, Ref.~\onlinecite{fernandez2024} developed a Monte Carlo numerical method to obtain more robust and accurate emitted electron energy spectra that includes additional IIEE processes such as potential emission or Auger electron emission.
Future work may consider interfacing the Monte Carlo results with the continuum kinetic method.

While we focused on the plasma-material interactions around Z-pinch electrodes, the IIEE model developed in this paper is sufficiently generalized so that it is applicable for any application with an ion bombardment of a metallic surface.
Furthermore, the results for the plasma interaction with biased electrodes can be applied to arcjets\cite{wollenhaupt2018overview} and tokamak edge biasing.\cite{weynants1993}

\section{Summary and Conclusions} \label{s:conclusions}

For plasma facing components near high energy ions, it is important to understand how ion induced electron emission (IIEE) affects the plasma and sheath formation.

We develop a general framework for performing continuum kinetic simulations with IIEE. 
An ionization cascade mechanism\cite{schou1980}, which is a function of the ion\cite{andersen1977,icru49_3} and electron\cite{nguyentruong2015} stopping powers, is used to model the emitted electron energy spectrum.
This model is applicable for proton, deuteron, or triton impacts on a metallic wall material.
A logarithmic Gaussian fit\cite{scholtz1996,bradshaw2024} is used to approximate the model for easier implementation as a Vlasov/Boltzmann equation boundary condition.\cite{bradshaw2024}
For some materials, such as copper, gold, and aluminum, the model secondary electron yield (SEY) agrees well with experimental data.

The IIEE model is used to better understand the plasma-material interactions near the electrodes of Z-pinch thermonuclear fusion reactors.
We perform 1X-1V unmagnetized simulations of a Z-pinch proton-electron plasma between two copper electrodes with bias potentials from 0 to \SI{9}{\kilo\volt} with and without IIEE. 
A classical sheath forms at both electrodes with the plasma potential being greater than both wall potentials.
The potential difference accelerates the ions to high energies which, upon impacting the electrode, induce a beam of emitted electrons.
These electrons are accelerated by the potential difference into the domain.
Due to the larger potential difference, more electrons are emitted at the cathode versus the anode.
The SEY increases with bias potential for both electrodes.
The electron emissions allow for a negative electron particle flux to exist at the cathode.
This causes an increase in current, assuming uniform cylindrical distribution in the Z-pinch, to \SI{273}{\kilo\ampere} at \SI{9}{\kilo\volt} with emissions compared to \SI{130}{\kilo\ampere} without emissions.
The experimental value of \SI{200}{\kilo\ampere} is most closely matched by the \SI{4}{\kilo\volt} case of \SI{201}{\kilo\ampere}.\cite{zhang2019sustained}
There are, however, other physics considerations such as the inclusion of a magnetic field, oxidation of the electrode, electron induced electron emission, and increased temperatures for scaled Z-pinch plasmas that will modify this result.
Nevertheless, it is clear that IIEE increases the current and helps explain the discrepancy between theory\cite{stangeby2000plasma} and experiment.\cite{zhang2019sustained}

At 0 and \SI{1}{\kilo\volt} bias potentials, the plasma potential is lower with emissions than without.
This is the expected result due to the emitted electrons reducing the positive space charge in the sheath.
Above \SI{1}{\kilo\volt}, however, the plasma potential increases for the IIEE cases.
The emitted electrons are accelerated to higher energies into the domain by the sheath potential difference.
Higher bias potentials accelerate the electrons to higher energies.
These high energy electrons collisionally heat the domain electrons resulting in an overall temperature increase with larger temperatures closer to the cathode.
A simple theoretical model, Eq.~\ref{eq:phip}, shows that the increase in electron temperature causes an increase in the plasma potential.
This increased potential further accelerates ions to higher energies at the electrodes resulting in larger SEY than predicted from simulations without emissions.

At a bias potential of \SI{8}{\kilo\volt}, the SEY becomes greater than 1. 
This would typically form a space-charge limited\cite{hobbs1967SCL,schwager1993SCL,mcadams2012} or inverse sheath.\cite{campanell2013negative,campanell2016}
Despite this, a classical sheath forms for all bias potentials tested because the space charge current density limit is larger than the emitted electron particle flux.
In addition, there is no backflow saturation\cite{campanell2025} of the current at the anode based on the collisionality of the plasma.

\begin{acknowledgments}
	This work was supported by the Department of Energy ARPA-E BETHE program under award number DE-AR0001263.   
	
	The authors acknowledge funding from Dr. Lindsay V. Goodwin's New Jersey Institute of Technology startup funds.
	
	This work was supported by the U.S. Department of Energy’s Scientific Discovery through Advanced Computing (SCIDAC) CEDA project and Distinguished Scientist Award program under contract number DE- AC02-09CH11466 for the Princeton Plasma Physics Laboratory.  
	
	The authors acknowledge Advanced Research Computing at Virginia Tech for providing computational resources and technical support that have contributed to the results reported within this paper. URL: \url{http://www.arc.vt.edu} 
	
	The authors acknowledge the Princeton Research Computing resources at Princeton University which is a consortium of groups led by the Princeton Institute for Computational Science and Engineering (PICSciE) and Office of Information Technology's Research Computing. 
	
	The authors acknowledge HYAK at the University of Washington for computing resources.
	
	The authors thank Dr. Hieu T. Nguyen-Truong for useful insights on calculating the low energy electron stopping powers.

\end{acknowledgments}

\section*{Data Availability Statement}
Readers may reproduce our results and also use Gkeyll for their applications. The code and input files used here are available online. 
Full installation instructions for Gkeyll are provided on the Gkeyll website.\cite{gkylDocs} The code can be installed on Unix-like 
operating systems (including Mac OS and Windows using the Windows Subsystem for Linux) either by installing the pre-built binaries using 
the conda package manager (\url{https://www.anaconda.com}) or building the code via sources. The input files used here are under version 
control and can be obtained from the repository at \url{https://github.com/ammarhakim/gkyl-paper-inp/tree/master/2025_PoP_IIEESheaths}.

\bibliography{reference}

\end{document}